\PassOptionsToPackage{unicode}{hyperref}
\PassOptionsToPackage{hyphens}{url}
\PassOptionsToPackage{dvipsnames,svgnames,x11names}{xcolor}
\documentclass[
  12pt]{article}
\usepackage{amsmath,amssymb}
\usepackage{array}
\usepackage{graphicx}
\usepackage{adjustbox}
\usepackage{float}
\usepackage{enumerate}
\usepackage{multirow}
\usepackage{multicol}
\usepackage{bbm}
\usepackage{iftex}
\ifPDFTeX
  \usepackage[T1]{fontenc}
  \usepackage[utf8]{inputenc}
  \usepackage{textcomp} 
\fi
\usepackage{lmodern}
\ifPDFTeX
\fi
\IfFileExists{upquote.sty}{\usepackage{upquote}}{}
\IfFileExists{microtype.sty}{
  \usepackage[]{microtype}
  \UseMicrotypeSet[protrusion]{basicmath} 
}{}
\makeatletter
\@ifundefined{KOMAClassName}{
  \IfFileExists{parskip.sty}{%
    \usepackage{parskip}
  }{
    \setlength{\parindent}{0pt}
    \setlength{\parskip}{6pt plus 2pt minus 1pt}}
}{
  \KOMAoptions{parskip=half}}
\makeatother
\usepackage{xcolor}
\makeatletter
\ifx\paragraph\undefined\else
  \let\oldparagraph\paragraph
  \renewcommand{\paragraph}{
    \@ifstar
      \xxxParagraphStar
      \xxxParagraphNoStar
  }
  \newcommand{\xxxParagraphStar}[1]{\oldparagraph*{#1}\mbox{}}
  \newcommand{\xxxParagraphNoStar}[1]{\oldparagraph{#1}\mbox{}}
\fi
\ifx\subparagraph\undefined\else
  \let\oldsubparagraph\subparagraph
  \renewcommand{\subparagraph}{
    \@ifstar
      \xxxSubParagraphStar
      \xxxSubParagraphNoStar
  }
  \newcommand{\xxxSubParagraphStar}[1]{\oldsubparagraph*{#1}\mbox{}}
  \newcommand{\xxxSubParagraphNoStar}[1]{\oldsubparagraph{#1}\mbox{}}
\fi
\makeatother

\usepackage{longtable,booktabs,array}
\usepackage{calc} 
\usepackage{etoolbox}
\makeatletter
\patchcmd\longtable{\par}{\if@noskipsec\mbox{}\fi\par}{}{}
\makeatother
\IfFileExists{footnotehyper.sty}{\usepackage{footnotehyper}}{\usepackage{footnote}}
\makesavenoteenv{longtable}
\usepackage{graphicx}
\makeatletter
\def\fps@figure{htbp}
\makeatother

\makeatletter
\@ifpackageloaded{caption}{}{\usepackage{caption}}
\AtBeginDocument{%
\ifdefined\contentsname
  \renewcommand*\contentsname{Table of contents}
\else
  \newcommand\contentsname{Table of contents}
\fi
\ifdefined\listfigurename
  \renewcommand*\listfigurename{List of Figures}
\else
  \newcommand\listfigurename{List of Figures}
\fi
\ifdefined\listtablename
  \renewcommand*\listtablename{List of Tables}
\else
  \newcommand\listtablename{List of Tables}
\fi
\ifdefined\figurename
  \renewcommand*\figurename{Figure}
\else
  \newcommand\figurename{Figure}
\fi
\ifdefined\tablename
  \renewcommand*\tablename{Table}
\else
  \newcommand\tablename{Table}
\fi
}
\@ifpackageloaded{float}{}{\usepackage{float}}
\floatstyle{ruled}
\@ifundefined{c@chapter}{\newfloat{codelisting}{h}{lop}}{\newfloat{codelisting}{h}{lop}[chapter]}
\floatname{codelisting}{Listing}

\makeatother
\makeatletter
\@ifpackageloaded{caption}{}{\usepackage{caption}}
\@ifpackageloaded{subcaption}{}{\usepackage{subcaption}}
\makeatother

\usepackage[]{natbib}
\usepackage{bookmark}

\IfFileExists{xurl.sty}{\usepackage{xurl}}{} 
\hypersetup{
  pdftitle={Title},
  pdfauthor={Author 1; Author 2},
  pdfkeywords={3 to 6 keywords, that do not appear in the title},
  colorlinks=true,
  linkcolor={blue},
  filecolor={Maroon},
  citecolor={Blue},
  urlcolor={Blue},
  pdfcreator={LaTeX via pandoc}}

\usepackage{authblk}

\newcommand{\anon}{1}

\begin{document}

\def\spacingset#1{\renewcommand{\baselinestretch}%
{#1}\small\normalsize} \spacingset{1}


\if1\anon
{
  \title{\bf A Bayesian Geoadditive Model for Spatial Disaggregation}

\date{}

\author[1]{Sara Rutten
\footnote{Corresponding author. {\textit{E-mail address}: sara.rutten@uhasselt.be}}}
\author[1,2]{Thomas Neyens}
\author[1]{Elisa Duarte}
\author[1]{Christel Faes}

\affil[1]{Interuniversity Institute for Biostatistics and statistical Bioinformatics (I-BioStat), Data Science Institute (DSI), Hasselt University, Hasselt, Belgium}
\affil[2]{L-BioStat, Department of Public Health and Primary Care, KU Leuven, Leuven, Belgium}

  \maketitle
} \fi

\if0\anon
{
  \bigskip
  \bigskip
  \bigskip
  \begin{center}
    {\LARGE\bf Title}
\end{center}
  \medskip
} \fi

\bigskip
\begin{abstract}
We present a novel Bayesian spatial disaggregation model for count data, providing fast and flexible inference at high resolution. First, it incorporates non-linear covariate effects using penalized splines, a flexible approach that is not typically included in existing spatial disaggregation methods. Additionally, it employs a spline-based low-rank kriging approximation for modeling spatial dependencies. The use of Laplace approximation provides computational advantages over traditional Markov Chain Monte Carlo (MCMC) approaches, facilitating scalability to large datasets. 
We explore two estimation strategies: one using the exact likelihood and another leveraging a spatially discrete approximation for enhanced computational efficiency. Simulation studies demonstrate that both methods perform well, with the approximate method offering significant computational gains. We illustrate the applicability of our model by disaggregating disease rates in the United Kingdom and Belgium, showcasing its potential for generating high-resolution risk maps. By combining flexibility in covariate modeling, computational efficiency and ease of implementation, our approach offers a practical and effective framework for spatial disaggregation.
\end{abstract}

\noindent%
\vfill

\newpage
\spacingset{1.8} 

\section{Introduction}
Developing high-resolution risk maps is an important task in many scientific disciplines, including epidemiology, ecology and environmental science. However, the response variable of interest is often only available as aggregated data over larger defined areas. Therefore, spatial disaggregation models recently became a popular area of research as these models aim to describe the variation of the variable of interest within aggregated areas, allowing for the construction of fine-scale maps \citep{Pittiglio2018, Brus2018, Li2012, Keil2013}. 

\cite{Diggle2013} introduced a Bayesian model to disaggregate spatial disease count data, based on an underlying log-Gaussian Cox process (LGCP). This approach can be used to combine data across multiple spatial scales, meaning that covariate data at a smaller spatial scale than the response data can be integrated. However, the method employs a data augmentation technique, resulting in high computation times. To deal with this issue, \cite{Johnson2019} introduced a spatially discrete approximation to the LGCP model (SDALGCP). Although this approach speeds up the estimation process significantly, the use of the Monte Carlo Markov chain (MCMC) algorithm still results in relatively high computation times. The SDALGCP model is implemented in the eponymous Rpackage \texttt{SDALGCP}.

Another approach is provided by a class of non-Bayesian spatial disaggregation methods using the idea of composite link models (CLM), proposed by \cite{Thompson1981}. CLMs are extensions of generalized linear models, designed to link each observation to multiple predicted values. Assuming an underlying continuous spatial process that can be modeled as a high-resolution square lattice, the areal count can be viewed as the sum of the counts of the lattice cells within the area \citep{Nandi2023}. Hence, the structure of a composite link model can be retrieved. This idea has been applied for disaggregation purposes by \cite{Ayma2016}, using the penalized composite link model approach (PCLM) of \cite{Ellers2007} together with a mixed model strategy. This model has been extended to a spatial temporal context by \cite{Lee2022}. Both models rely on a penalized spline evaluated over the spatial coordinates to capture the spatial process and therefore do not allow obtaining estimates of the underlying spatial correlation structure. 

A solution to this problem comes in the form of the classical geostatistical kriging approach. Area-to-point (ATP) kriging has been introduced by \cite{Kyriakidis2004}. To solve the ATP kriging systems, a point-support semivariogram is needed. This function cannot be calculated directly from areal data but should be derived from a `regularized' experimental semivariogram \citep{Goovaerts2006}. Over the years, several methods have been proposed to address this difficulty \citep{Pardo-Igzquiza2007, Goovaerts2008, Nagle2011, Truong2014}. The approach of \cite{Goovaerts2008} has been implemented in an R-package called \texttt{atakrig} \citep{Hu2020}. However, this method only serves the purpose of disaggregating, not allowing for the estimation of covariate effects. Although, this R-package is tailored to continuous data with a Gaussian assumption, Rcode to perform ATP Poisson kriging is available on \href{https://github.com/DavidPayares/ATA-Poisson-Cokriging/tree/main}{GitHub} \citep{payares2025}. 

A Bayesian kriging approach that allows for the inclusion of covariates and avoids the pre-specification of a semivariogram, has been introduced by \cite{Nandi2023}. Following the idea of a composite link model, their approach assumes that areal counts can be obtained by summing the counts of all lattice cells within an area. In contrast to the models introduced by \cite{Ayma2016} and \cite{Lee2022}, they include a spatial component that is assumed to follow a Gaussian process. For computational speed, Laplace approximation is used for model estimation. To deal with the potentially high computational burden arising from the inversion of the spatial covariance matrix, the \texttt{disaggregation} R package approximates Gaussian random fields (GRFs) using the Stochastic Partial Differential Equation (SPDE) framework \citep{Lindgren2011}. This requires the construction of a mesh, which can be technically challenging. In the special case of Gaussian data and a linear link function, Integrated Nested Laplace Approximation (INLA) \citep{rue2009}, through the \href{http://www.r-inla.org}{\texttt{R-INLA}} package, can also be used to fit these models \citep{Moraga2017}, but it lacks flexibility in the case of other distributions. Unlike the composite link models, these kriging models provide an estimate of the underlying spatial correlation structure but they do not allow the flexible modeling of smooth covariates.

Another promising solution to the computational burden associated with the estimation of the Gaussian process used in kriging, comes from a low-rank approximation. Spline-based approaches have been proposed in geostatistical data analysis to flexibly model the spatial component, without the need for mesh construction. For example, \cite{Kammann2003} introduced a geoadditive framework that combines a spline-based low-rank kriging approach with additive models that allow the inclusion of non-linear covariate relationships.  It is important to note that, although a spline based approach is used, this low-rank approximation is linked to the underlying spatial correlation structure, allowing to obtain the relevant estimates. This geostatistical approach has been implemented in a Bayesian framework by \cite{Sumalinab2024}. They use Laplace approximation to reduce computation time, compared to the traditional MCMC approaches. However, to the best of our knowledge, it has not been used for disaggregation purposes. 

Despite the effectiveness of existing methods \citep{Johnson2019, Ayma2016, Nandi2023}, there remains a need for a spatial disaggregation technique that combines the estimation of the spatial correlation structure with flexible modeling of smooth covariate effects and is computationally fast. In this paper, we introduce a novel spatial disaggregation model for count data, using a spline-based low-rank kriging approximation and incorporating smooth covariate effects by the use of P-splines. Similar to the method of \cite{Nandi2023}, this modeling approach can be linked to the idea of composite link models. By leveraging Laplace approximations, the proposed method efficiently estimates the posterior distribution of the regression coefficients, offering computational advantages compared to MCMC approaches \citep{gressani2018, lambert2023, Sumalinab2024, gressani2024}.

In Section~2, we introduce the methodology behind the new spatial disaggregation method. In Section~3, a simulation study is conducted to investigate the performance of the model. Section~4 contains two applications of the method on a dataset containing mortality counts in Belgium and a dataset on primary biliary cirrhosis (PBC) incidence in Newcastle upon Tyne, UK.

\section{Methodology}
Consider a location $\boldsymbol{w}$ with coordinates $\boldsymbol{w} = (w_{1},w_{2})$. Assume the true (but latent) count at $\boldsymbol{w}$ is given by $y(\boldsymbol{w})$. Let $m(\boldsymbol{w})$ and $r(\boldsymbol{w})$ represent the population size (e.g., the population at risk) and intensity (e.g., the disease's incidence rate) at location $\boldsymbol{w}$, respectively. Then $y(\boldsymbol{w})$ can be modeled as a realization of a Poisson variable with mean $\mu(\boldsymbol{w}) = m(\boldsymbol{w})r(\boldsymbol{w})$, such that $E(y(\boldsymbol{w})) = \mu(\boldsymbol{w})$. Now, let $Y_i$ represent the actual reported count in area $R_i \ (i =1, \ldots, n)$. \cite{Diggle2013} argue that it is natural to treat these counts as aggregated values, meaning that the average number of cases in area $R_i$ can be expressed as:
\begin{align}
    \label{eq:exact_integral1}
    \mu_i &=\int_{R_i}{m(\boldsymbol{w})r(\boldsymbol{w}) d\boldsymbol{w} }.  
\end{align}

\subsection{Continuous geoadditive model}
A geostatistical model for the underlying intensity $r(\boldsymbol{w})$ can be formulated. \cite{Sumalinab2024} present a Bayesian P-splines based approach to geoadditive modeling, combining the flexibility of a generalized additive model with spatial smoothing. Their model can be formulated as follows:
\begin{align}
    \label{eq:geostat_model}
    \log(r(\boldsymbol{w})) = \beta_0 + \sum_{k=1}^p\beta_kx_{k}(\boldsymbol{w}) + \sum_{j=1}^q f_j(z_{j}(\boldsymbol{w}))+  s(\boldsymbol{w}).
\end{align}
The covariates $x_1 \ldots x_p$ are linear predictors whereas the covariates $z_1, \ldots, z_q$ are included as non-linear functions. The non-linear functions are formulated as a P-spline:
\begin{align*}
    f_j(z_{j}(\boldsymbol{w})) = \sum_{k=1}^{K}{\theta_{jk}b_{jk}(z_{j}(\boldsymbol{w}))}, \ \ j=1,\ldots,q,      
\end{align*}
in which $\boldsymbol{\theta}_j = (\theta_{j1}, \theta_{j2}, \ldots, \theta_{jK})^T$ follows a Gaussian prior $(\boldsymbol{\theta}_j \vert \lambda_j) \sim \mathcal{N}_K(\boldsymbol{0},(\lambda_j\boldsymbol{P})^{-1})$ with penalty parameter $\lambda_j$ and matrix $\boldsymbol{P}$ derived from a difference matrix \citep{lang2004}.

The spatial term $s(\boldsymbol{w})$ accounts for spatial correlation and is modeled through a classical kriging approach. To address the high computational cost of kriging, \cite{Sumalinab2024} employ a low-dimensional representation of the spatial component, building on the approach of \cite{Kammann2003}, who use the kriging covariance function as basis functions.  Commonly used correlation functions are:
\begin{itemize}
    \itemsep0em 
    \item Exponential: $R_\rho(\boldsymbol{d}) = \exp(-\rho \lVert\boldsymbol{d}\rVert),$
    \item Matérn with $\nu = 3/2$: $R_\rho(\boldsymbol{d}) = \exp(-\rho \lVert\boldsymbol{d}\rVert)(1+\rho\lVert \boldsymbol{d} \rVert),$
    \item Spherical: $R_\rho(\boldsymbol{d}) = (1-1.5\rho \lVert \boldsymbol{d} \rVert + 0.5 \rho^3 \lVert \boldsymbol{d} \rVert^3) \mathbbm{1}(\lVert \boldsymbol{d} \rVert \leq \rho^{-1}),$
    \item Circular: $R_\rho(\boldsymbol{d}) = (1-\frac{2}{\pi}(\vartheta\sqrt{1-\vartheta^2} + \arcsin{\vartheta}))$ with $\vartheta = \min(\rho \lVert \boldsymbol{d} \rVert,1),$
\end{itemize}
where the parameter $\rho$ represents the range parameter used in kriging and $\lVert \cdot \rVert$ refers to the Euclidean distance. Note that $ \nu=3/2$ for the Matérn function corresponds to the simplest Matérn function that results in differentiable surface estimates \citep{Kammann2003}.
The spatial component can then be modeled as,
\begin{align}
    \label{eq:smooth term}
    s(\boldsymbol{w}) = \beta_{w1} w_{1}+\beta_{w2}w_{2}+ \sum_{s=1}^{S}{\phi_{s}(\rho)u_s}+\epsilon(\boldsymbol{w}),
\end{align}
where $\phi_{s}(\rho) = R_{\rho}(\boldsymbol{w}-\boldsymbol{\kappa}_s)$ are the basis functions and the coefficients $\boldsymbol{u} = (u_1, \ldots u_S)^T$ are assumed to follow a multivariate normal distribution $(\boldsymbol{u} \vert \lambda_{spat},\rho) \sim \mathcal{N}_{S}(0,(\lambda_{spat}\boldsymbol{\Omega_\rho})^{-1})$ where $\lambda_{\text{spat}}>0$. The $S \times S$ matrix $\boldsymbol{\Omega_\rho} = R_\rho(\boldsymbol{\kappa}_s - \boldsymbol{\kappa}_{s'})$ is constructed using a subset of spatial coordinates $\boldsymbol{\kappa}_s \ (s=1,\ldots,S)$. Following \cite{Sumalinab2024}, these knots can be chosen through the use of a space-filling algorithm \citep{Johnson1990, nychka1998}. The error term $\epsilon$ is included to account for inaccuracies arising from the low-rank approximation. These errors are assumed to follow a normal distribution with mean $0$ and constant variance $\sigma^2$.

\subsection{A spatial disaggregation model}
\begin{figure}[h]
    \centering
    \includegraphics[width=0.8\linewidth]{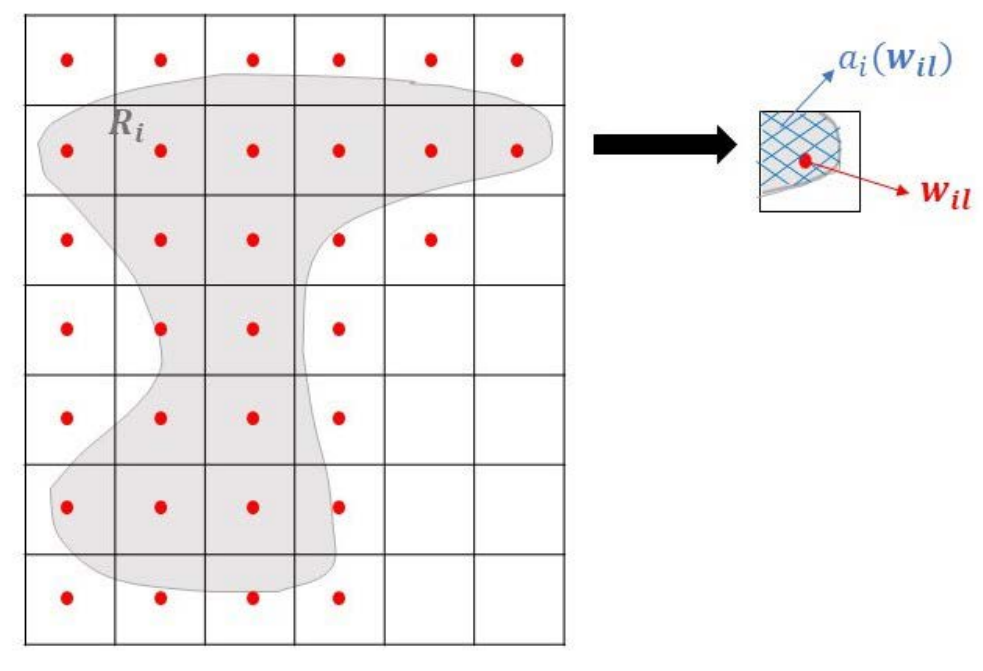}
    \caption{Visualization of the fine grid approach: in red, $\boldsymbol{w_{il}} \ (l=1,\ldots,L_i)$ denote the centers of the grid cells intersecting $R_i$ and in blue, $a_i(\boldsymbol{w_{il}})$ denotes the percentage of a grid cell covered by $R_i$. }
    \label{fig:Visualisation_grid}
\end{figure}

Combining equation \eqref{eq:exact_integral1} and the geoadditive model in equation \eqref{eq:geostat_model}, we can assume that the reported counts $Y_i$ are, conditional on $s(\boldsymbol{w})$, mutually independent Poisson variables with means:
\begin{align}
 \label{eq:exact_integral2} 
    \mu_i = \int_{R_i}{m(\boldsymbol{w})\exp\left(\beta_0 + \sum_{k=1}^p\beta_kx_{k}(\boldsymbol{w}) + \sum_{j=1}^q f_j(z_{j}(\boldsymbol{w}))+  s(\boldsymbol{w})\right)d\boldsymbol{w} }.
\end{align}
We can approximate integral~\eqref{eq:exact_integral2} on a fine grid. Denote by $\boldsymbol{w_{il}} \ (l=1,\ldots,L_i)$ the $L_i$ centers of the grid cells that intersect area $R_i$ and let $a_i(\boldsymbol{w_{il}})$ represent the percentage of grid cell $l$, centered in $ \boldsymbol{w_{il}}$, that is covered by $R_i$ (Figure \ref{fig:Visualisation_grid}). 

Then the approximation of integral~\eqref{eq:exact_integral2} becomes:
\begin{align}
    \label{eq:approx_integral}
    \mu_i &\approx\sum_{l=1}^{L_i}{a_i(\boldsymbol{w_{il}}) m(\boldsymbol{w_{il}})\exp\left(\beta_0 + \sum_{k=1}^p\beta_kx_{k}(\boldsymbol{w_{il}}) + \sum_{j=1}^q f_j(z_{j}(\boldsymbol{w_{il}}))+  s(\boldsymbol{w_{il}})\right)}.
\end{align}
Following equation~\eqref{eq:smooth term}, we model the smooth term as:
\begin{align*}
    s(\boldsymbol{w_{il}}) = \beta_{w1} w_{1il}+\beta_{w2}w_{2il}+ \sum_{s=1}^{S}{\phi_{ils}(\rho)u_s}+\epsilon_i,
\end{align*}
with $\phi_{ils}(\rho) = R_\rho(\boldsymbol{w_{il}}-\boldsymbol{\kappa}_s)$. Since the response data are available at the areal level, we include an area-specific independent random effect $\epsilon_i$, representing a population-weighted average of the error terms at the fine grid, i.e., 
\begin{align*}
    \epsilon_i = \frac{\sum_{l=1}^{L_i}{a_i(\boldsymbol{w_{il}})m(\boldsymbol{w_{il}})\epsilon(\boldsymbol{w_{il}})}}{\sum_{l=1}^{L_i}{a_i(\boldsymbol{w_{il}})m(\boldsymbol{w_{il}})}}.
\end{align*}
Therefore, they are modeled as $\epsilon_i \sim \mathcal{N}\biggl(0,\frac{\sum_{l=1}^{L_i}{(a_i(\boldsymbol{w_{il}})m_i(\boldsymbol{w_{il}}))^2}}{\bigl(\sum_{l=1}^{L_i}{a_i(\boldsymbol{w_{il}})m_i(\boldsymbol{w_{il}})}\bigl)^2}\sigma^2 \biggl)$.

In matrix notation, the model can be formulated as:
\begin{align*}
    \boldsymbol{\mu} = \boldsymbol{A_1}\exp(\boldsymbol{X} \boldsymbol{\beta} + \sum_{j=1}^{q}{\boldsymbol{B_j(z_j)}\boldsymbol{\theta_j}}+\boldsymbol{\Phi(\rho)} \boldsymbol{u} + \boldsymbol{E}\boldsymbol{\epsilon}),
\end{align*}
where $\boldsymbol{\mu} = (\mu_1, \ldots, \mu_n)^T$. Let $N = \sum_{i=1}^{n}{L_i}$. Then $X$ is a design matrix of dimension $N \times (p+3)$, where $\boldsymbol{X} = (\boldsymbol{X_1}^T, \ldots \boldsymbol{X_n}^T)^T$ with the l-th row of $L_i \times (p+3)$ dimensional matrix $\boldsymbol{X_i}$ given by $(1,x_1(\boldsymbol{w_{il}}), \ldots, x_p(\boldsymbol{w_{il}}), w_{1il}, w_{2il})$ and coefficient vector $\boldsymbol{\beta} = (\beta_0, \beta_1, \ldots, \beta_p,\beta_{w1}, \beta_{w2})^T$. Similarly, $\boldsymbol{B_j(z_j)}$ is an $N \times K$ matrix $\boldsymbol{B_j(z_j)}=(\boldsymbol{B_{j,1}(z_{j})}^T,\ldots,\boldsymbol{B_{j,n}(z_{j})}^T)^T$ and the l-th row of $L_i \times K$ dimensional matrix $\boldsymbol{B_{j,i}(z_{j})}$ is equal to $ \biggl(b_{j1}(z_{j}(\boldsymbol{w_{il}})), \ldots, b_{jK}(z_{j}(\boldsymbol{w_{il}}))\biggl)$. The coefficient vector $\boldsymbol{\theta_j} = (\theta_{j1}, \theta_{j2}, \ldots, \theta_{jK})^T$. Also, $\boldsymbol{\Phi(\rho)}$ is a $N \times S$ matrix with $\boldsymbol{\Phi(\rho)}=(\boldsymbol{\Phi_1(\rho)}^T,\ldots,\boldsymbol{\Phi_n(\rho)}^T)^T$ and the l-th row of $L_i \times S$ dimensional matrix $\boldsymbol{\Phi_i(\rho)}$ is equal to $(\phi_{il1}(\rho), \phi_{il2}(\rho), \ldots, \phi_{ilS}(\rho))$. The coefficient vector $\boldsymbol{u}$ is equal to $\boldsymbol{u} = (u_1, u_2, \ldots, u_S)^T$ and $\boldsymbol{\epsilon}$ is a $n$-dimensional vector $(\epsilon_1, \ldots \epsilon_n)^T$. Matrix $\boldsymbol{E}$ is a $N \times n$ design matrix with column $i$ equal to
\begin{align*}
(\underbrace{0, \ldots, 0}_{\sum_{l=1}^{i-1}{L_l} \text{ times}}, \underbrace{1, \ldots, 1}_{L_i \text{ times}}, \underbrace{0, \ldots, 0}_{\sum_{l=i+1}^{n}{L_l} \text{ times}})^T.
\end{align*}
Finally, $\boldsymbol{A_1}$ is a weight-matrix of dimension $n \times N$ where row $i$ is defined as 
\begin{align*}
    (\underbrace{0, \ldots, 0}_{\sum_{l=1}^{i-1}{L_l} \text{ times}}, a_i(\boldsymbol{w_{i1}})m_i(\boldsymbol{w_{i1}}), \ldots, a_i(\boldsymbol{w_{iL_i}})m_i(\boldsymbol{w_{iL_i}}), \underbrace{0, \ldots, 0}_{\sum_{l=i+1}^{n}{L_l} \text{ times}}).
\end{align*}

Priors are defined for all regression parameters. We specify $\boldsymbol{\beta} \sim \mathcal{N}(0,\boldsymbol{V_\beta}^{-1})$ with $\boldsymbol{V_\beta} = \zeta \boldsymbol{I}$ and $\zeta$ small. Furthermore, denote the global design matrix by $\boldsymbol{C_\rho} = [\boldsymbol{X}: \boldsymbol{B_1(z_1)}: \boldsymbol{B_2(z_2)} : \ldots : \boldsymbol{B_q(z_q)} : \boldsymbol{\Phi(\rho)}:\boldsymbol{E}]$ and the parameter vector $\boldsymbol{\xi} = (\boldsymbol{\beta}^T, \boldsymbol{\theta_1}^T, \ldots, \boldsymbol{\theta_q}^T, \boldsymbol{u}^T,\boldsymbol{\epsilon}^T)^T$. Let $\lambda_{q+1}:=\lambda_{spat}$, $\lambda_{q+2} := 1/\sigma^2$ and hence $\boldsymbol{\lambda} = (\lambda_1, \ldots \lambda_q, \lambda_{q+1},\lambda_{q+2})^T$. 

Denote the precision matrix of $\boldsymbol{\xi}$ by $\boldsymbol{Q_{\xi}^{\lambda}} = \text{blkdiag}(\boldsymbol{V_\beta}, \lambda_1\boldsymbol{P_1}, \ldots \lambda_q\boldsymbol{P_q}, \lambda_{q+1}\boldsymbol{\Omega_{\rho}},\lambda_{q+2}\boldsymbol{G})$ where $\lambda_{q+2}\boldsymbol{G}$ is the precision matrix of the vector $\boldsymbol{\epsilon}$. Hence, $\boldsymbol{G}$ is a diagonal matrix with diagonal entry $i$ equal to $$\frac{(\sum_{l=1}^{L_i}{a_i(\boldsymbol{w_{il}})}m_i(\boldsymbol{w_{il}}))^2}{\sum_{l=1}^{L_i}{(a_i(\boldsymbol{w_{il}})m_i(\boldsymbol{w_{il}}))^2}}.$$ 

The Bayesian model is given by:
\begin{align*}
   &(y_i \vert  \boldsymbol{\xi}) \sim \text{Poisson}(\mu_i) \text{ with } \boldsymbol{\mu} = \boldsymbol{A_1} \exp(\boldsymbol{C_\rho} \boldsymbol{\xi}), \\
   &(\boldsymbol{\xi} \vert \boldsymbol{\lambda},\rho) \sim \mathcal{N}(0,(\boldsymbol{Q_{\xi}^{\lambda}})^{-1}), \\
   &(\lambda_j \vert \delta_j) \sim \mathcal{G}\left(\frac{\nu}{2}, \frac{\nu \delta_j}{2}\right) \ \ j=1,\ldots,q+2, \\
   &\delta_j \sim \mathcal{G}(a_\delta,b_\delta) \ \ j=1,\ldots,q+2, \\
   &p(\rho) \propto \mathcal{G}\left(\frac{\nu}{2}, \frac{\nu \delta_\rho}{2}\right),  \\
   &\delta_\rho \sim \mathcal{G}(a_\delta,b_\delta),
\end{align*}
where $\mathcal{G}(a,b)$ is a Gamma distribution with mean $a/b$ and variance $a/b^2$. This prior specification is selected based on the work of \cite{jullion2007}, who suggest that this choice is robust if $a=b$ sufficiently small (e.g. $10^{-5}$) and fixed $\nu$ (e.g. $\nu=3$ in this paper).

Note that $\mu_i$ in equation~\eqref{eq:approx_integral} is constructed as a summation of exponential terms, deviating from the typical log-link structure commonly used in Poisson regression. However, \cite{Johnson2019} proposed a spatially discrete approximation method that allows for approximating integral~\eqref{eq:exact_integral1} in a way that restores the standard log-link structure. This approach will be detailed as well and both methods will be compared in the simulation study presented in Section~\ref{sec:Simulation_study}.

\subsubsection{A spatially discrete approximation}
\label{subsec:SDA}
Similarly as done by \cite{Johnson2019}, we can simplify the computation of integral~\eqref{eq:exact_integral1} by using a spatially discrete approximation of the log expected incidence $\log(r(\boldsymbol{w}))$, replacing $\log(r(\boldsymbol{w}))$ for every $\boldsymbol{w} \in R_i$ with the weighted average within region $R_i$. This method is theoretically founded on a first-order Taylor approximation of $r(\boldsymbol{w})$ (see Supplementary Materials for more details). We can then approximate $\log(r(\boldsymbol{w}))$ in region $R_i$ by:
\begin{align}
\label{eq:SPDA_mu}
    \log(r(\boldsymbol{w})) &\approx \int_{R_i}{v_i(\boldsymbol{w}) \left(\beta_0 + \sum_{k=1}^p\beta_kx_{k}(\boldsymbol{w}) + \sum_{l=1}^q f_l(z_{l}(\boldsymbol{w}))+  s(\boldsymbol{w})\right)d\boldsymbol{w}} \nonumber \\
    & = \beta_0 + \sum_{k=1}^p\beta_kx_{ki}^* + \sum_{l=1}^q f^*_{li}+  s^*_i,
\end{align}
with $v_i(\boldsymbol{w}) = m(\boldsymbol{w})/m_i$ and $m_i = \int_{R_i}{m(\boldsymbol{w})d\boldsymbol{w}}$, reflecting the potential nonhomogeneous distribution of disease cases within $R_i$. If the continuous distribution $m(\boldsymbol{w})$ is unavailable, we can alternatively set  $v_i(\boldsymbol{w})=1/ \vert R_i\vert$. Further, $\beta_j$ is the regression coefficient for the aggregate explanatory variable $x_{ji}^* \ (j =1,\ldots p)$,  $f_{ji}^* \ (j=1,\ldots, q)$ is the aggregated smoothing function and $s_i^*$ the aggregated spatial term. 

Combining \eqref{eq:exact_integral1} and \eqref{eq:SPDA_mu}, the expected counts in region $R_i$ become: 
\begin{align}
\label{eq:approx_integral_2}
    \mu_i &\approx \int_{R_i}{m(\boldsymbol{w})\exp\left(\beta_0 + \sum_{k=1}^p\beta_kx_{ki}^* + \sum_{j=1}^q f^*_{ji}+  s^*_i\right)d\boldsymbol{w}} \nonumber \\
    & = m_i \exp\left(\beta_0 + \sum_{k=1}^p\beta_kx_{ki}^* + \sum_{j=1}^q f^*_{ji}+  s^*_i\right),
\end{align}
which again resembles the log-link structure commonly used in Poisson regression.

To ease notation, we denote by $\widetilde{g(\boldsymbol{w_i})}$ the approximation of the integral $\int_{R_i}{v_i(\boldsymbol{w}) g(\boldsymbol{w})d\boldsymbol{w}}$, defined as:
$$\widetilde{g(\boldsymbol{w_i})}=\frac{\sum_{l=1}^{L_i}{v_i(\boldsymbol{w_{il}})a_i(\boldsymbol{w_{il}})g(\boldsymbol{w_{il}})}}{\sum_{l=1}^{L_i}{v_i(\boldsymbol{w_{il}})a_i(\boldsymbol{w_{il}})}}.$$
The aggregated spatial term $s_i^*$ can then be written as:
\begin{align*}
    s^*_i &= \widetilde{s(\boldsymbol{w_i})}= \beta_{w1}\widetilde{w_{1i}} + \beta_{w2}\widetilde{w_{2i}}  + \sum_{s=1}^{S}u_s \widetilde{\phi_{is}(\rho)} 
    + \widetilde{\epsilon(\boldsymbol{w_i})}.
\end{align*}
The terms $\left(\epsilon_1^* \ldots \epsilon_n^*\right)^T=\left(\widetilde{\epsilon(\boldsymbol{w_1})},\ldots,\widetilde{\epsilon(\boldsymbol{w_n})}\right)^T$ are again multivariate Gaussian with mean $0$ and the variance of $\epsilon_i^*$ equal to: $$\frac{\sum_{l=1}^{L_i}{(v_i(\boldsymbol{w_{il}})a_i(\boldsymbol{w_{il}}))^2\sigma^2}}{(\sum_{l=1}^{L_i}{v_i(\boldsymbol{w_{il}})a_i(\boldsymbol{w_{il}})})^2}.$$

Similarly, the non-linear covariate functions $f^*_{ji}$ can be written as:
\begin{align*}
    f_{ji}^* &= \widetilde{f_j(z_j(\boldsymbol{w_i}))} 
    = \sum_{k=1}^{K}{\theta_{jk} \widetilde{b_{jk}(z_j(\boldsymbol{w_i}))}}.
\end{align*}
and the linear covariate as $x_{ki}^* = \widetilde{x_k(\boldsymbol{w_i})}$.

In matrix notation, the model can be formulated as:
\begin{align}
\label{eq:approx_matrix}
    \log(\boldsymbol{\mu}) = \boldsymbol{A_2}\biggl(\boldsymbol{X}\boldsymbol{\beta} + \sum_{j=1}^{q}{\boldsymbol{B_j(z_j)}\boldsymbol{\theta}_j} + \boldsymbol{\Phi(\rho)} \boldsymbol{u} + \boldsymbol{E}\boldsymbol{\epsilon^*}\biggl)+\log(m_i),
\end{align}
where $\boldsymbol{X}$, $\boldsymbol{E}$, $\boldsymbol{B_j(z_j)}$ and $\boldsymbol{\Phi(\rho)}$ are the continuous design matrices of dimensions $N \times (p+3)$, $N \times n$, $N \times K$ and $N \times S$ respectively, as defined earlier. Row $i$ of weight-matrix $\boldsymbol{A_2}$ of dimension $n \times N$ can now be defined as: 
\begin{align*}
    (\underbrace{0, \ldots, 0}_{\sum_{l=1}^{i-1}{L_l} \text{ times}}, \frac{v_i(\boldsymbol{w_{i1}})a_i(\boldsymbol{w_{i1}})}{\sum_{l=1}^{L_i}{v_i(\boldsymbol{w_{il}})a_i(\boldsymbol{w_{il}})}}, \ldots, \frac{v_i(\boldsymbol{w_{iL_i}})a_i(\boldsymbol{w_{iL_i}})}{\sum_{l=1}^{L_i}{v_i(\boldsymbol{w_{il}})a_i(\boldsymbol{w_{il}})}}, \underbrace{0, \ldots, 0}_{\sum_{l=i+1}^{n}{L_l} \text{ times}}).
\end{align*}

We assume similar priors as before. Denote the global design matrix by $\boldsymbol{C_\rho} = [\boldsymbol{A_2X}: \boldsymbol{A_2B_1}: \boldsymbol{A_2B_2} : \ldots : \boldsymbol{A_2B_q} : \boldsymbol{A_2\Phi(\rho)}:\boldsymbol{A_2E}]$ and the full parameter vector $\boldsymbol{\xi} = (\boldsymbol{\beta}^T, \boldsymbol{\theta_1}^T, \ldots, \boldsymbol{\theta_q}^T, \boldsymbol{u}^T,\boldsymbol{\epsilon^*}^T)^T$. The matrix $\boldsymbol{G}$ is now a diagonal matrix with diagonal entry $i$ equal to $$\frac{(\sum_{l=1}^{L_i}{v_i(\boldsymbol{w_{il}})a_i(\boldsymbol{w_{il}})})^2}{\sum_{l=1}^{L_i}{(v_i(\boldsymbol{w_{il}})a_i(\boldsymbol{w_{il}}))^2}}.$$ 
The Bayesian model is then given by:
\begin{align*}
   &(y_i \vert  \boldsymbol{\xi}) \sim \text{Poisson}(m_ir_i) \text{ with } \log(\boldsymbol{r}) = \boldsymbol{C_\rho} \boldsymbol{\xi}, \\
   &(\boldsymbol{\xi} \vert \boldsymbol{\lambda},\rho) \sim \mathcal{N}(0,(\boldsymbol{Q_{\xi}^{\lambda}})^{-1}), \\
   &(\lambda_j \vert \delta_j) \sim \mathcal{G}\left(\frac{\nu}{2}, \frac{\nu \delta_j}{2}\right) \ \ j=1,\ldots,q+2, \\
   &\delta_j \sim \mathcal{G}(a_\delta,b_\delta) \ \ j=1,\ldots,q+2, \\
   &p(\rho) \propto \mathcal{G}\left(\frac{\nu}{2}, \frac{\nu \delta_\rho}{2}\right),  \\
   &\delta_\rho \sim \mathcal{G}(a_\delta,b_\delta).
\end{align*}

\subsection{Laplace approximation}
For both estimation methods, the posterior of $\boldsymbol{\xi}$ conditional on the penalty vector $\boldsymbol{\lambda}$ and $\rho$, can be denoted by:
\begin{align*}
    p(\boldsymbol{\xi} \vert \boldsymbol{\lambda},\rho; \mathcal{D}) & \propto \mathcal{L}(\boldsymbol{\xi}, \rho;\mathcal{D}) p(\boldsymbol{\xi} \vert \boldsymbol{\lambda}, \rho) \\
    & \propto \exp{\left( \sum_{i=1}^{n}(y_i\log(\mu_i) - \exp(\mu_i)\bigl)\right)}\exp{\left(-\frac{1}{2}(\boldsymbol{\xi}'\boldsymbol{Q_\xi^\lambda} \boldsymbol{\xi} \bigl)\right)} \\
    & = \exp{\left( \sum_{i=1}^{n}(y_i\log(\mu_i) - \exp(\mu_i)-\frac{1}{2}(\boldsymbol{\xi}'\boldsymbol{Q_\xi^\lambda} \boldsymbol{\xi} \bigl)\right)}, 
\end{align*}
with $\mu_i$ equal to \eqref{eq:approx_integral} and \eqref{eq:approx_integral_2} for the exact and approximate method, respectively.
The Laplace approximation for the conditional posterior of $\boldsymbol{\xi}$ is $\tilde{p}_G(\boldsymbol{\xi} \vert \boldsymbol{\lambda}, \rho; \mathcal{D})=\mathcal{N}(\hat{\boldsymbol{\xi}}_\lambda,\hat{\boldsymbol{\Sigma}}_\lambda)$ where $\hat{\boldsymbol{\xi}}_\lambda$ is the posterior mode and $\hat{\boldsymbol{\Sigma}}_\lambda$ corresponds to the inverse of the negative Hessian matrix evaluated at the posterior mode. We can now derive the joint posterior of the hyperparameters $\boldsymbol{\lambda}, \rho$ and $\boldsymbol{\delta}$ similar to \cite{Sumalinab2024}. This function is then maximized to obtain an a posteriori estimate for $\boldsymbol{v}= (\log{(\lambda_1)}, \ldots, \log{(\lambda_{q+2})})^T$ and $v_\rho = \log{(\rho)}$. Details can be found in the Supplementary materials. 

\section{Simulation study}
\label{sec:Simulation_study}
We conducted a simulation study to assess the performance of the proposed estimation method and compared its performance against three existing disaggregation methods, namely the disaggregation approach of \cite{Nandi2023}, implemented in the \texttt{disaggregation} R-package, ATP Poisson kriging \citep{payares2025} and SDALGCP \citep{Johnson2019}. We also compared it with a discrete disease mapping method, namely the BYM model \citep{besag1991}. For general understanding, we provide a brief overview of these methods, summarizing the key principles.
\subsubsection*{The method of the \texttt{disaggregation} package}
Similar to our approach, the method of \cite{Nandi2023} assumes that the observed (aggregated) counts can be considered realizations of a Poisson variable with mean given by equation \eqref{eq:exact_integral1}. The underlying intensity $r(\boldsymbol{w})$ for $\boldsymbol{w} \in R_i$ is modeled by:
\begin{align*}
    \log(r(\boldsymbol{w})) = \beta_0 + \sum_{k=1}^{p}{\beta_k x_k(\boldsymbol{w})+s(\boldsymbol{w})+u_i},
\end{align*}
where $u_i$ is a region-specific i.i.d. Gaussian random effect and $s(\boldsymbol{w})$ a Gaussian random field. The Gaussian random field is modeled using a Matérn covariance function with $\nu = 1$. The approximation of integral \eqref{eq:exact_integral1} then becomes:
\begin{align*}
    \mu_i \approx \sum_{l=1}^{N_i}{m(\boldsymbol{w_{il}})\exp\biggl(\beta_0 + \sum_{k=1}^{p}{\beta_k x_k(\boldsymbol{w})+s(\boldsymbol{w})+u_i\biggl)}},
\end{align*}
with $N_i$ the number of grid cells for which the centre lies within $R_i$ (i.e. $N_i \leq L_i$). The Gaussian random field is approximated with a Gaussian Markov random field (GMRF) and solved using the stochastic partial differential equation (SPDE) approach \citep{Lindgren2011}, which requires the construction of a mesh.

\subsubsection*{ATP Poisson kriging}
Following \cite{Goovaerts2006}, the ATP kriging estimator for the rate $r(\boldsymbol{w})$ with $\boldsymbol{w} \in R_i$ is estimated as a linear combination of the observed rate $Y_i/m_i$ and the rates of $(K-1)$ neighboring units with $K \leq N$:
\begin{align*}
    r(\boldsymbol{w}) = \sum_{j=1}^{K}{\lambda_i(\boldsymbol{w})\frac{Y_j}{m_j}}.
\end{align*}
In this equation $\lambda_j(\boldsymbol{w})$ represents the weight assigned to rate $Y_j/m_j$ when estimating the incidence at location $\boldsymbol{w}$. These kriging weights are obtained by solving a system of linear equations, constraining $\sum_{j=1}^{K}{\lambda_j(\boldsymbol{w})} = 1$. To solve this system, a model for the point-support spatial covariance function is required. A method for deriving such a model from areal data has been proposed by \cite{Goovaerts2008}.

\subsubsection*{The SDALGCP method}
This method proposed by \cite{Johnson2019} again starts from the idea represented by integral \eqref{eq:exact_integral1}. However, the intensity $r(\boldsymbol{w})$ is now assumed to follow the model:
\begin{align*}
    \log(r(\boldsymbol{w})) = \beta_0 + \sum_{k=1}^{p}{\beta_k x_k(\boldsymbol{w})} + s(\boldsymbol{w}),
\end{align*}
with Gaussian process $s(\boldsymbol{w})$. Using a spatially discrete approximation, the area-level expected counts can be approximated by:
\begin{align*}
    \mu_i \approx m_i \exp\biggl(\beta_0 + \sum_{k=1}^{p}\beta_kx_{ki}^*+s_i^*\biggl),
\end{align*}
with $x_{ki}^*$ the aggregated linear covariates and $s^* = (s_1^*, \ldots s_N^*)$ a multivariate Gaussian, with an exponential covariance function by default. As this method assumes an underlying Log-Gaussian Cox process instead of a geostatistical model, a suitable number of locations is drawn in every region $R_i$, according to a class of inhibition processes \citep{Diggle2013_book}. The aggregated covariates $x_{ki}^*$ and spatial process $s_i^*$ are computed by a weighted average over the sampled points, rather than across all grid cells, as is done by our method and by the Bayesian spatial disaggregation method in the \texttt{disaggregation} package. The MCMC method is used to obtain parameter estimates.

\subsubsection*{The BYM model}
Following the BYM model \citep{besag1991}, the observed counts $Y_i$ are mutually independent Poisson variables, conditional on a Gaussian process $S_i$, with expected counts modeled by:
\begin{align*}
    \mu_i = m_i \exp\biggl(\beta_0 + \sum_{k=1}^{p}\beta_kx_{ki}+S_i\biggl),
\end{align*}
where $x_{ki}$ is the area level covariate and $S_i$ can be written as $S_i = S_{1i}+S_{2i}$. The process $S_{1i}$ is then modeled as:
\begin{align*}
    (S_{1i} \vert S_{1h}, j \neq h) \sim \mathcal{N}\biggl(n_j^{-1}\sum_{j \sim h}{S_{1h}}, (\tau_1 n_j)^{-1}\biggl),
\end{align*}
where $j \sim h$ denotes that regions $R_j$ and $R_h$ are neighbors and $n_j$ represents the number of neighbors of region $R_j$. The independent process $S_{2i}$ can be modeled as $S_{2i} \sim \mathcal{N}(0,\tau_2)$.

\subsection{Simulation set-up}
We simulated the spatial effect over a $100 \times 100$ square as a Gaussian random field with variance $\sigma^2 = 0.7$, reflecting a moderate level of spatial variability, and a Matérn covariance function with $\nu = 1$, the default option in the disaggregation method of \cite{Nandi2023}. To consider both small and large spatial correlation, the range parameter $(1/\rho)$ was chosen equal to $3$ and $10$ in scenario (a) and (b) respectively. For both scenarios, we defined the simulation surface as:
\begin{align*}
    \log(r(\boldsymbol{w})) = \beta_0 + \beta_1 x_1(\boldsymbol{w})+s(\boldsymbol{w}),
\end{align*}
with $\beta_0 = -3, \ \beta_1 = -1.5$, and with $x_1$ simulated to take values approximately within the range $(0,1.5)$. Note that we did not include a smooth effect to enable a fair comparison with other disaggregation methods that do not allow for the estimation of flexible non-linear trends.

To explore the ability of the methods to recover the underlying smooth surface, we considered three different configurations where the size of the areas progressively increases: $4\times 4$ (Area 1), $10 \times 10$ (Area 2) and $20 \times 20$ (Area 3). We then simulated a dataset of area-level counts from a Poisson process with mean given by equation~\eqref{eq:exact_integral1}.

\begin{figure}
    \centering
    \begin{minipage}{.25\textwidth}
        \centering
        \includegraphics[width=\linewidth]{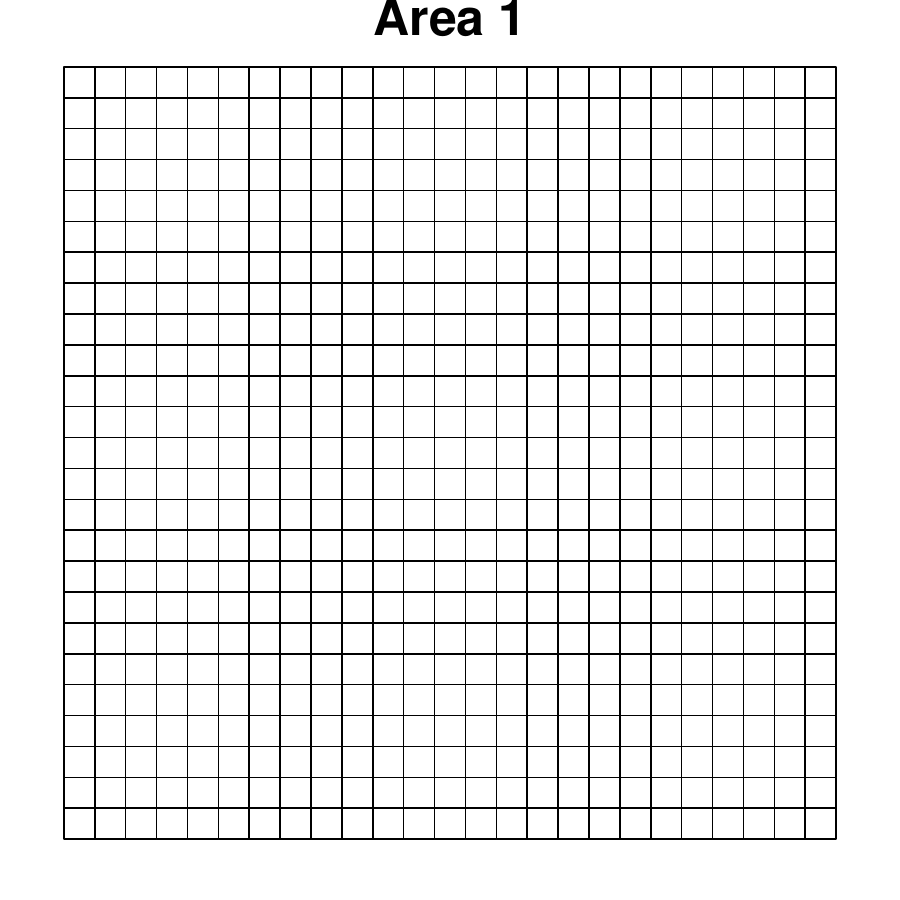}
    \end{minipage}%
    \begin{minipage}{0.25\textwidth}
        \centering
        \includegraphics[width=\linewidth]{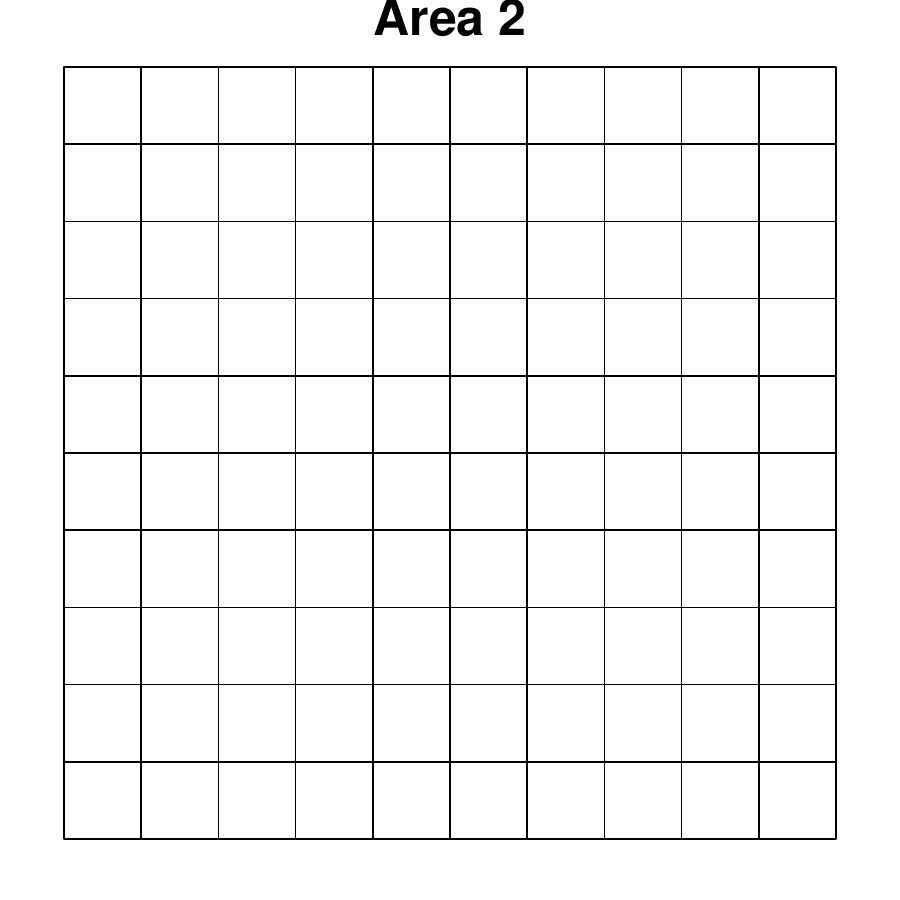}
    \end{minipage}
        \begin{minipage}{0.25\textwidth}
        \centering
        \includegraphics[width=\linewidth]{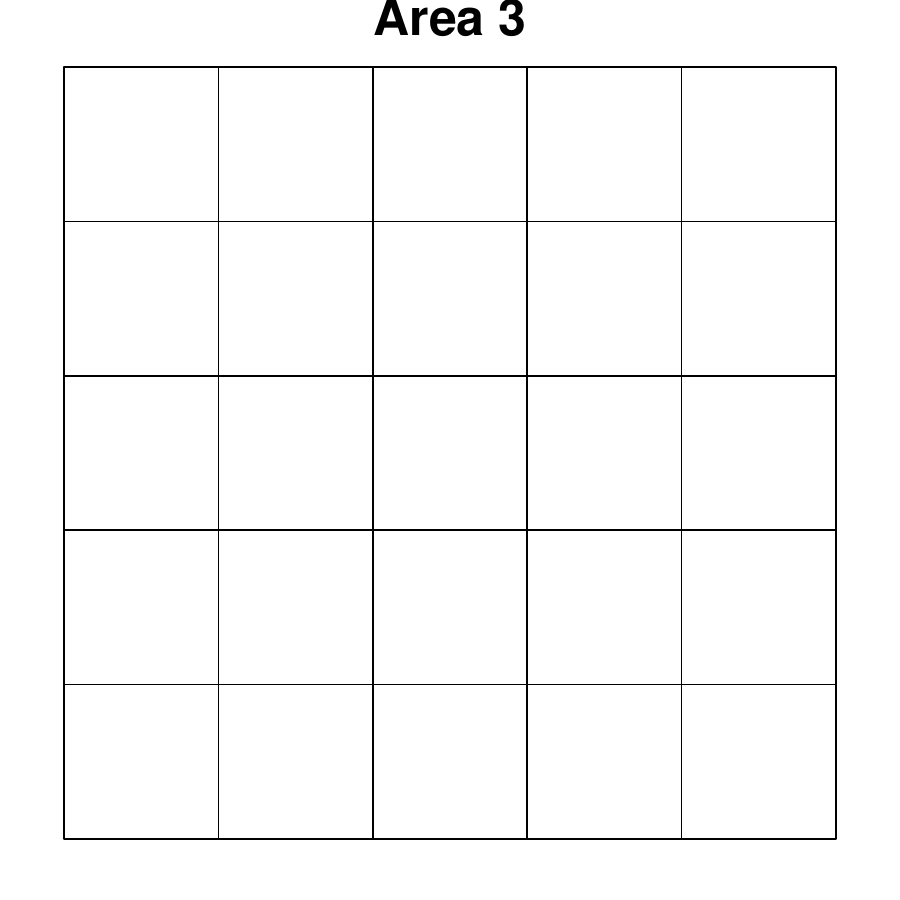}
    \end{minipage}
        \caption{The three different area configurations: Area 1 ($625$ areas of size $4 \times 4$), Area 2 ($100$ areas of size $10 \times 10$) and Area 3 ($25$ areas of size $20 \times 20$).}
        \label{fig:raster_configuration}
\end{figure}

We performed $100$ simulations for every combination of scenario (range equal to $3$ or $10$) and area configuration. We investigated the performance of our estimation method based on the exact likelihood, which we coin the Spline Spatial Disaggregation Exact Method (henceforth, SSDEM), and the approximation method proposed in Section~\ref{subsec:SDA}, called the Spline Spatial Disaggregation Approximation Method (henceforth, SSDAM) using a Matérn covariance function ($\nu = 3/2$). We used $350$, $200$ and $50$ spatial knots $\boldsymbol{\kappa}_S$ for area configuration 1, 2 and 3, respectively. The disaggregation grid resolution is chosen to be $1 \times 1$. We evaluated both methods in terms of the RMSE and coverage of the spatial effect and intensity at the grid cell level as well as the average count at the area level. The point estimates are defined as the posterior median, while credible intervals are constructed using the $2.5\%$ and $97.5\%$ percentile of the posterior distribution. More information can be found in the Supplementary Materials. We compared the RMSE and coverage of our methods with results obtained from the \texttt{disaggregation} package, the SDALGCP method and ATP Poisson kriging, when available. 

\subsection{Simulation results}
Results for scenario (a) and (b), respectively mild and strong spatial correlation, are given in Table \ref{tab:simulation}. A graphical representation of the RMSEs of the different methods can be found in the Supplementary Materials (Figure S1 and Figure S2). It is important to note that the SDALGCP method failed in $45\%$ of the simulations for scenario (b) with area configuration 1, due to singularity issues. In contrast, none of the other methods encountered errors. Consequently, the results for this specific setting are based on only 55 successful simulations. In the remaining settings, issues were minimal: failure rates were $2\%$ in scenario (b) with area configuration 2 and $3\%$ in scenario (a) with area configuration 1. All other configurations yielded a $100\%$ success rate across simulations.

\begin{figure}
    \centering
    \includegraphics[width=0.75\linewidth]{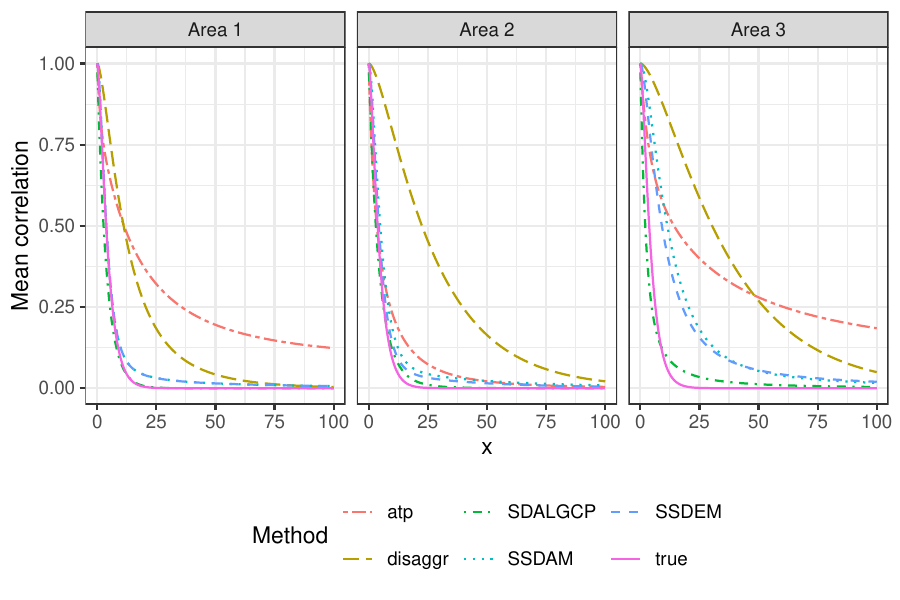}
    \caption{The mean estimated correlation function for a true range of $3$ (scenario a) with the different estimation methods, compared with the true correlation function.}
    \label{fig:cor_mean3}
\end{figure}

\begin{figure}
    \centering
    \includegraphics[width=0.75\linewidth]{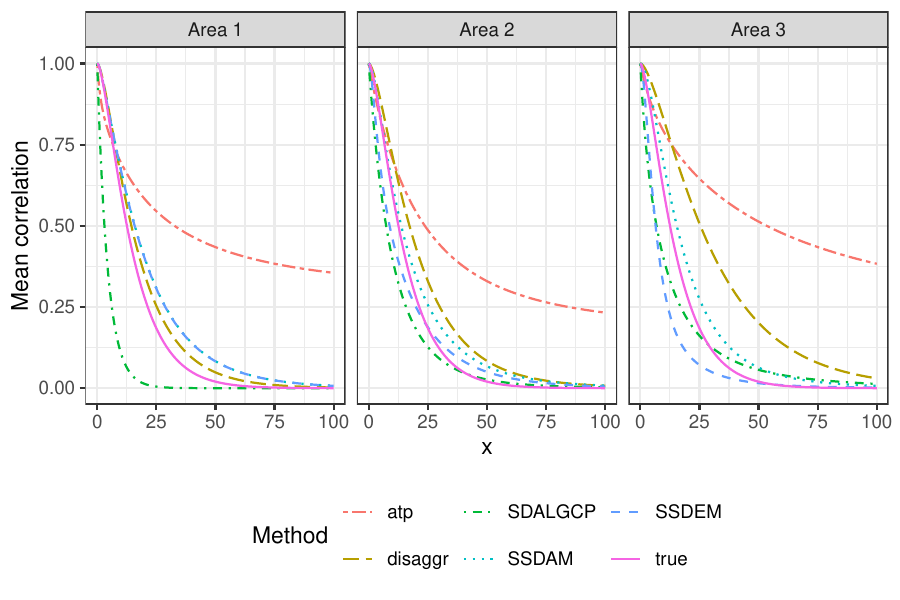}
    \caption{The mean estimated correlation function for a true range of $10$ (scenario b) with the different estimation methods, compared with the true correlation function.}
    \label{fig:cor_mean10}
\end{figure}

The results show that SSDEM and SSDAM perform similarly in terms of RMSE and coverage of the spatial term and continuous incidence, with the latter offering shorter computation times, especially with increasing sample size. As expected, the RMSE at the grid level increases with increasing area size, as less detailed information is available. Figures \ref{fig:cor_mean3} and \ref{fig:cor_mean10} show the estimated mean correlation functions in scenarios (a) and (b), respectively. Both the SSDEM and SSDAM methods recover the true correlation function well, particularly in area configurations 1 and 2.

Compared to these, the method of the \texttt{disaggregation} package performs similarly in scenario (b) but underperforms in scenario (a), showing higher RMSE and undercoverage of the spatial term for all area configurations. This is also confirmed by Figure \ref{fig:cor_mean3}, showing that the \texttt{disaggregation} package overestimates the spatial range for all three area configurations in scenario (a). 

The SDALGCP method, using $110,000$ iterations with a burn-in of $10,000$ samples and retaining every $10$th sample, yields comparable RMSEs of the grid-level spatial effect in scenario (a), although there is some undercoverage for area configurations 2 and 3. However, the results show increased RMSEs in scenario (b). Figure \ref{fig:cor_mean10} also shows that the SDALGCP method underestimates the spatial range for area configuration 3 in scenario (b). Finally, as previously noted, the SDALGCP method frequently encountered singularity issues in this specific setting. 

The ATP method, producing only disaggregated grid-level counts, yields RMSEs comparable to those of the other methods but exhibits some undercoverage in area configuration 3. Moreover, Figures \ref{fig:cor_mean3} and \ref{fig:cor_mean10} show that the correlation function is overestimated in almost all settings. A further examination of the results revealed some extremely high estimated ranges that substantially affected the mean. To address this, the median correlation estimates are also reported in the Supplementary Materials (Figures S3 and S4). While the median provides values closer to the true correlation function, the ATP method does still not outperform SSDEM or SSDAM in terms of closeness to the true values. 

We also compared the estimated spatial variance to the true variance of $0.7$ across all settings. As shown in Figure S5 and S6 (Supplementary Materials), all methods perform similarly, though ATP occasionally produces extreme outliers. 

At the area level, we can compare the SSDEM, SSDAM and SDALGCP disaggregation methods to the classical BYM method, fitted with INLA. The SSDAM and SDALGCP methods demonstrate RMSE performance comparable to that of the BYM model. For area configuration 3, the SSDEM method exhibits slightly higher RMSEs compared to the BYM method, likely due to the increased computational complexity of the model and the limited number of large areas available. However, for area configurations 1 and 2, its performance is comparable. The ATP and \texttt{disaggregation} package, in contrast, do not provide area-level estimates. 

Looking at the computation times, the BYM model consistently performs fastest; however, as it is not a disaggregation method, it does not offer any insights into the underlying smooth spatial effect or continuous incidence. Among the disaggregation methods, the \texttt{disaggregation} package is the fastest but shows lower performance than our proposed approach when the spatial range is small. Additionally, it does not produce area-level estimates. The ATP method also offers fast computation but lacks the ability to recover the underlying spatial surface and does not support the inclusion of covariates. The SSDAM method remains computationally efficient—typically completing within minutes—and its runtime can be further reduced through parallelization. It performs well across both small and large spatial ranges. The SSDEM method generally exhibits similar performance to SSDAM but requires longer computation times. Finally, the SDALGCP method is the most computationally intensive, due to MCMC sampling, and underperforms relative to our methods when the spatial range is large. It also does not provide estimates of the continuous incidence.

\renewcommand{\arraystretch}{0.6} 
\begin{table}
\caption{Simulation results for RMSE and Coverage for the spatial term, continuous incidence (Cont. inc.) and discrete average counts (Discr. inc.) under scenario (a) (range $=3$) and (b) (range $=10$). The estimation time is given in seconds. Results are given for the $6$ different estimation methods (SSDEM, SSDAM, disaggregation, SDALGCP, ATP, BYM).}
\label{tab:simulation}
\centering
\begin{adjustbox}{width=\linewidth} 
\begin{tabular}{cccccccccc}
\hline
\textbf{Scenario} & \textbf{Area configuration} &  \textbf{Estimation} & \textbf{time} & \multicolumn{2}{c}{\textbf{Spatial term}} & \multicolumn{2}{c}{\textbf{Cont. inc.}} & \multicolumn{2}{c}{\textbf{Discr. inc.}} \\ \cline{4-10}
                             & & \textbf{Method}   &  & RMSE & Cov  & RMSE & Cov & RMSE & Cov \\ \hline
 \multirow{18}{*}{(a) (Range $3$)} & \multirow{6}{*}{Area 1 ($4 \times 4$)} &  SSDEM  & 2066 & 0.71 & 0.97 & 0.03 & 0.96 & 0.31 & 0.91\\ 
                            & &SSDAM  & 356 & 0.71 & 0.97 & 0.03 & 0.95 & 0.31 & 0.92 \\ 
                            & &disaggregation  & 32 & 0.76 & 0.60 & 0.03 & 0.84 & $\cdot$ &  $\cdot$ \\  
                            & &SDALGCP  & 5513 & 0.71 & 0.98 & $\cdot$ & $\cdot$ & 0.32 & 0.97 \\  
                            & &ATP & 132 & $\cdot$ & $\cdot$ & 0.03 & 0.87 & $\cdot$ & $\cdot$\\
                            & &BYM & 17 & $\cdot$ & $\cdot$ & $\cdot$ & $\cdot$ & 0.33 & 0.94 \\
\cline{2-10}
 &\multirow{6}{*}{Area 2 ($10 \times 10$)} &  SSDEM  & 588 & 0.75 & 0.94 & 0.03 & 0.93 & 1.26 & 0.90\\ 
                            & &SSDAM  & 127 & 0.75 & 0.94 & 0.03 & 0.93 & 1.23 & 0.91 \\ 
                            & &disaggregation  & 32 & 0.80 & 0.45 & 0.03 & 0.65 & $\cdot$ &  $\cdot$ \\  
                            & &SDALGCP  & 1481 & 0.80 & 0.61 & $\cdot$ & $\cdot$ & 1.26 & 0.89 \\  
                            & &ATP & 143 & $\cdot$ & $\cdot$ & 0.03 & 0.88 & $\cdot$ & $\cdot$\\
                            & &BYM & 15 & $\cdot$ & $\cdot$ & $\cdot$ & $\cdot$ & 1.24 & 0.94 \\ 
\cline{2-10}
 &\multirow{6}{*}{Area 3 ($20 \times 20$)}  &  SSDEM  & 89 & 0.80 & 0.89 & 0.03 & 0.87 & 3.18 & 0.86\\ 
                            & &SSDAM  & 39 & 0.80 & 0.90 & 0.03 & 0.88 & 3.02 & 0.88 \\ 
                            & &disaggregation  & 32 & 0.82 & 0.34 & 0.03 & 0.41 & $\cdot$ &  $\cdot$ \\  
                            & &SDALGCP  & 1203 & 0.82 & 0.51& $\cdot$ & $\cdot$ & 3.02 & 0.84 \\  
                            & &ATP & 246 & $\cdot$ & $\cdot$ & 0.03 & 0.69 & $\cdot$ & $\cdot$\\
                            & &BYM & 15 & $\cdot$ & $\cdot$ & $\cdot$ & $\cdot$ & 2.90 & 0.94 \\
\hline

 \multirow{18}{*}{(b) (Range $10$)} & \multirow{6}{*}{Area 1 ($4 \times 4$)} &  SSDEM  & 2111 & 0.50 & 0.99 & 0.02 & 0.99 & 0.22 & 0.90\\ 
                            & &SSDAM  & 385 & 0.50 & 0.99 & 0.02 & 0.99 & 0.22 & 0.92 \\ 
                            & &disaggregation  & 32 & 0.51 & 0.91 & 0.02 & 0.87 & $\cdot$ &  $\cdot$ \\  
                            & &SDALGCP  & 5542 & 0.55 & 1.00 & $\cdot$ & $\cdot$ & 0.25 & 0.99 \\  
                            & &ATP & 134 & $\cdot$ & $\cdot$ & 0.02 & 0.92 & $\cdot$ & $\cdot$\\
                            & &BYM & 17 & $\cdot$ & $\cdot$ & $\cdot$ & $\cdot$ & 0.24 & 0.98 \\
\cline{2-10}
 &\multirow{6}{*}{Area 2 ($10 \times 10$)} &  SSDEM  & 605 & 0.54 & 0.99 & 0.02 & 0.99 & 1.29 & 0.90\\ 
                            & &SSDAM  & 131 & 0.52 & 0.99 & 0.02 & 0.99 & 1.15 & 0.93 \\ 
                            & &disaggregation  & 32 & 0.54 & 0.91 & 0.02 & 0.84 & $\cdot$ &  $\cdot$ \\  
                            & &SDALGCP  & 1507 & 0.61 & 0.87 & $\cdot$ & $\cdot$ & 1.16 & 0.94 \\  
                            & &ATP & 122 & $\cdot$ & $\cdot$ & 0.02 & 0.91 & $\cdot$ & $\cdot$\\
                            & &BYM & 15 & $\cdot$ & $\cdot$ & $\cdot$ & $\cdot$ & 1.24 & 0.9 \\ 
\cline{2-10}
 &\multirow{6}{*}{Area 3 ($20 \times 20$)}  &  SSDEM  & 104 & 0.61 & 0.97 & 0.02 & 0.96 & 3.64 & 0.89\\ 
                            & &SSDAM  & 42 & 0.59 & 0.98 & 0.02 & 0.97 & 3.16 & 0.91 \\ 
                            & &disaggregation  & 31 & 0.64 & 0.76 & 0.02 & 0.70 & $\cdot$ &  $\cdot$ \\  
                            & &SDALGCP  & 1254 & 0.68 & 0.80& $\cdot$ & $\cdot$ & 3.20 & 0.91 \\  
                            & &ATP & 229 & $\cdot$ & $\cdot$ & 0.02 & 0.84 & $\cdot$ & $\cdot$\\
                            & &BYM & 14 & $\cdot$ & $\cdot$ & $\cdot$ & $\cdot$ & 3.18 & 0.94 \\
\hline

\end{tabular}
\end{adjustbox}
\end{table}

\section{Data application}
We apply our method to two real-life datasets. Firstly, we analyze a dataset on primary biliary cirrhosis (PBC) incidence in Newcastle upon Tyne, UK. In this data application, we compare the obtained results using our method with different existing disaggregation methods. Our second data application focuses on mortality rates in Belgium. While the data are inherently fine-grained, we can aggregate them to a lower resolution, such as the municipality level. Disaggregating these data back to the higher resolution enables us to directly test whether our model successfully reconstructs detailed insights from aggregated data. This application provides a real-world, non-simulated setting to evaluate the disaggregation process. We make use of $350$ spatial knots $\boldsymbol{\kappa}_s$ and an exponential covariance function for both data applications. 

\subsection{Mapping of primary biliary cirrhosis risk}
This dataset contains incidence data on PBC at the LSOA level in Newcastle upon Tyne, UK. The dataset is freely available from the \texttt{SDALGCP} package in R. It contains PBC rates between 1987 and 1994 as well as the index of multiple deprivation (IMD) for the $545$ LSOAs. Since we include our covariates at the grid level, we extracted the IMD value at every grid cell, constructing a spatial raster. These data have previously been analyzed by \cite{Johnson2019}. Similar to their model, we included IMD as the only linear covariate. We compared the SSDEM method with the population-weighted SSDAM (SSDAM~I) and unweighted SSDAM (SSDAM~II) estimation methods as well as with the SDALGCP method of \cite{Johnson2019}, using the population-weighted version, area-to-point Poisson kriging and the estimation method implemented in the \texttt{disaggregation} package. To investigate the possible advantage of a non-linear effect, we also fitted the model, using the SSDAM~I method, allowing for a smooth effect of IMD (with $df$ = 10). Moreover, we compared the area-level estimates with the classical BYM model \citep{besag1991}. For the SDALGCP method, we used $1 100 000$ iterations of the MCMC algorithm with a burn-in of $100 000$ samples, retaining every $100$th sample, using the code provided by \cite{Johnson2019}. 

\begin{figure}
     \centering
     \begin{subfigure}{0.4\textwidth}
         \centering
         \includegraphics[width=\textwidth]{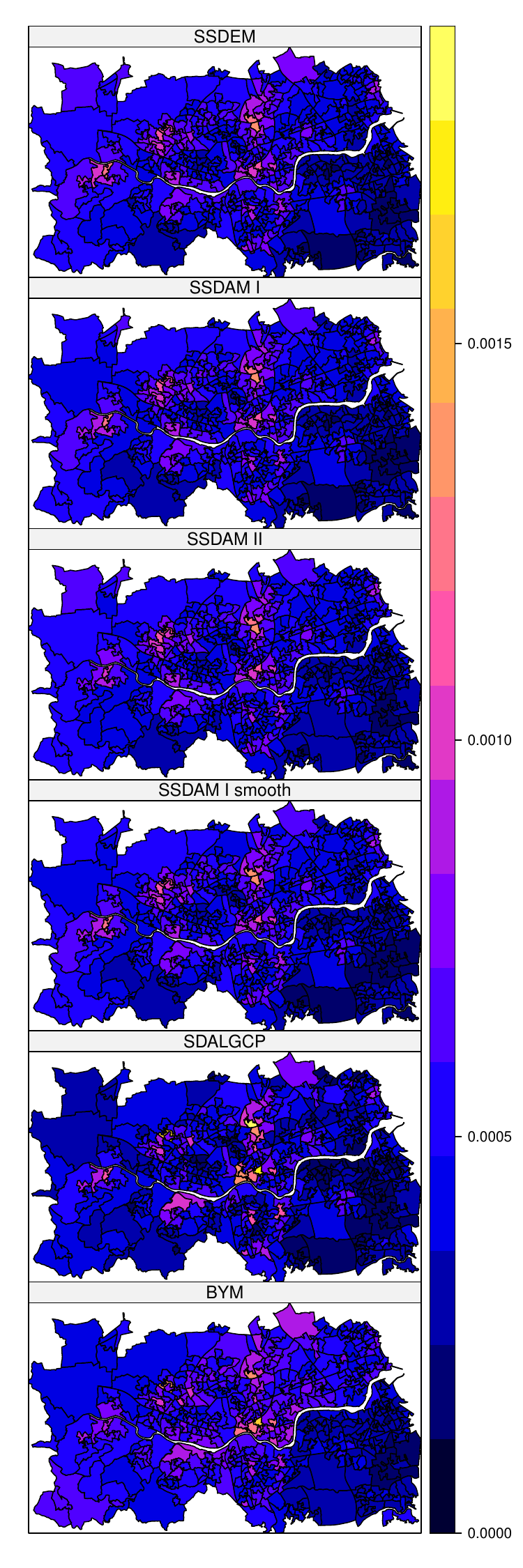}
         \caption{}
         \label{fig:incidence_uk}
     \end{subfigure}
     \hfill
     \begin{subfigure}{0.4\textwidth}
         \centering
         \includegraphics[width=\textwidth]{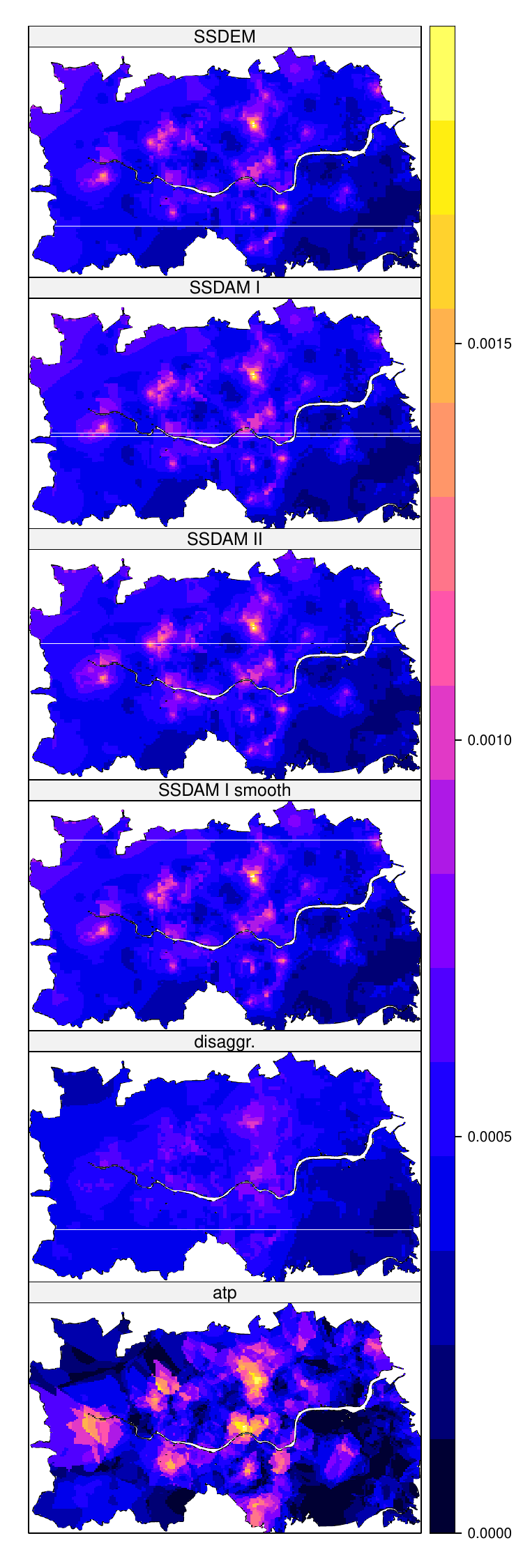}
         \caption{}
         \label{fig:spatial_uk}
     \end{subfigure}
    \caption{(a) Maps of the estimated biliary cirrhosis incidence in each LSOA of Newcastle upon Tyne, UK, from: SSDEM, SSDAM I, SSDAM II, SSDAM I smooth, SDALGCP and BYM. (b) Map of the continuous incidence from: SSDEM, SSDAM I, SSDAM II, SSDAM I smooth, disaggregation and ATP.}
    \label{fig:random_effect}
\end{figure}

\begin{figure}
    \centering
    \includegraphics[width=0.8\linewidth]{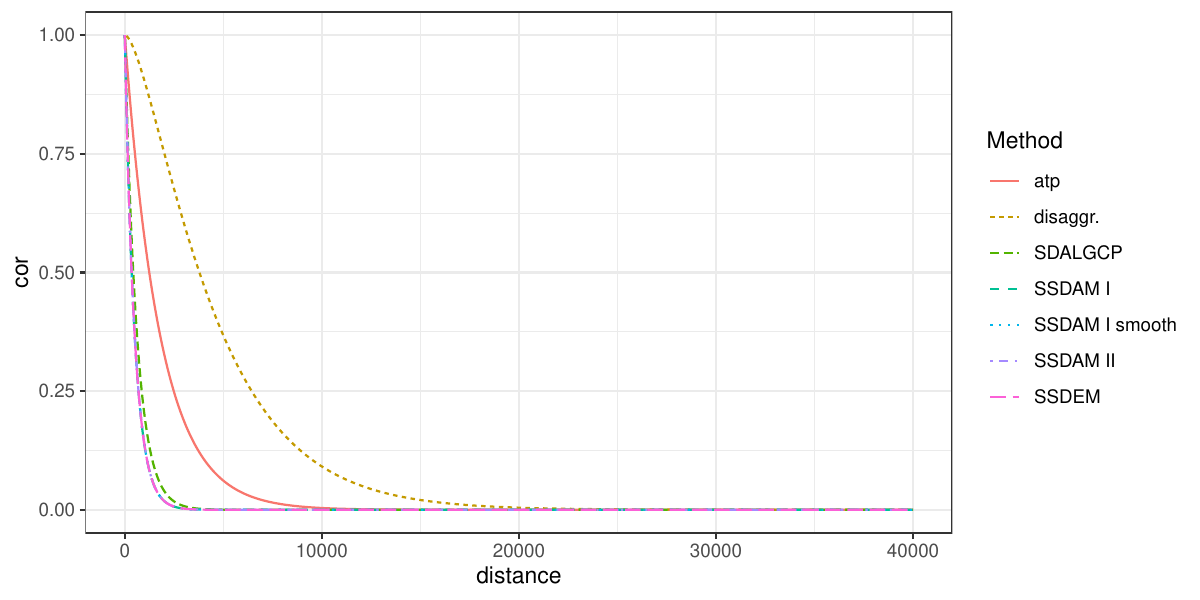}
    \caption{Estimated spatial covariance function from the disaggregation model (red), SDALGCP (brown), SSDAM I (green), SSDAM II (light blue), SSDAM II smooth(dark blue) and SSDEM (pink). }
    \label{fig:cov_uk}
\end{figure}

Figure \ref{fig:incidence_uk} shows a map of the estimated PBC incidence at the LSOA level from all models except \texttt{disaggregation} and ATP, which do not produce estimates at the area level. Similarly, an estimated map of the continuous incidence, for all methods except SDALGCP and BYM, can be found in Figure \ref{fig:spatial_uk}. The Supplementary Materials also contains a figure comparing the estimated spatial surface $\exp(s(x))$ (Figure S7). Furthermore, Figure \ref{fig:cov_uk} shows the estimated spatial correlation function from all models assuming an underlying continuous surface (i.e. all models except the BYM model). 

It can be seen that the estimated correlation function found by the SSDEM, linear SSDAM~I, smooth SSDAM~I, SSDAM~II and SDALGCP method are highly similar, in contrast to the correlation function found by the \texttt{disaggregation} package.

Correlations of the estimated area-level incidence, the estimated spatial effect $\exp(s(x))$ and estimated continuous incidence are also calculated. A high similarity between the SSDEM, SSDAM~I (linear and smooth) and SSDAM~II methods (correlations above $0.98$) is found. There is a relatively high correlation between our methods and the estimates from the SDALGCP model (correlation of approximately $0.87$ and $0.70$ for the area-level incidence and spatial effect respectively) and the estimates of the \texttt{disaggregation} package (correlation of approximately $0.76$ and $0.72$ for the grid-level incidence and spatial effect respectively). We find a correlation of approximately $0.85$ with the BYM model, at the area-level and a correlation of $0.63$ with the ATP model at the continuous incidence level. A detailed overview of the correlations can be found in the Supplementary Materials (Figure S8-S10). To investigate the sensitivity of the chosen covariance structure, we provided a comparison with the Matérn, circular and spherical correlation function, using the (linear) SSDAM~I and SSDEM method. This analysis revealed that this choice has minimal impact on the results (Figure S11 and S12 in the Supplementary Materials).

When allowing for a smooth relationship, the effect of IMD is estimated to be increasing but slightly non-linear, although credible intervals are large (see Figure S13 Supplementary Materials). We also provided a comparison of the estimated posterior distributions of the IMD effect for all models assuming a linear effect in the Supplementary Materials (Figure S14).

\subsection{Mapping of mortality rates}
\begin{figure}
    \centering
    \begin{subfigure}[b]{0.4\textwidth}
        \includegraphics[width=\linewidth]{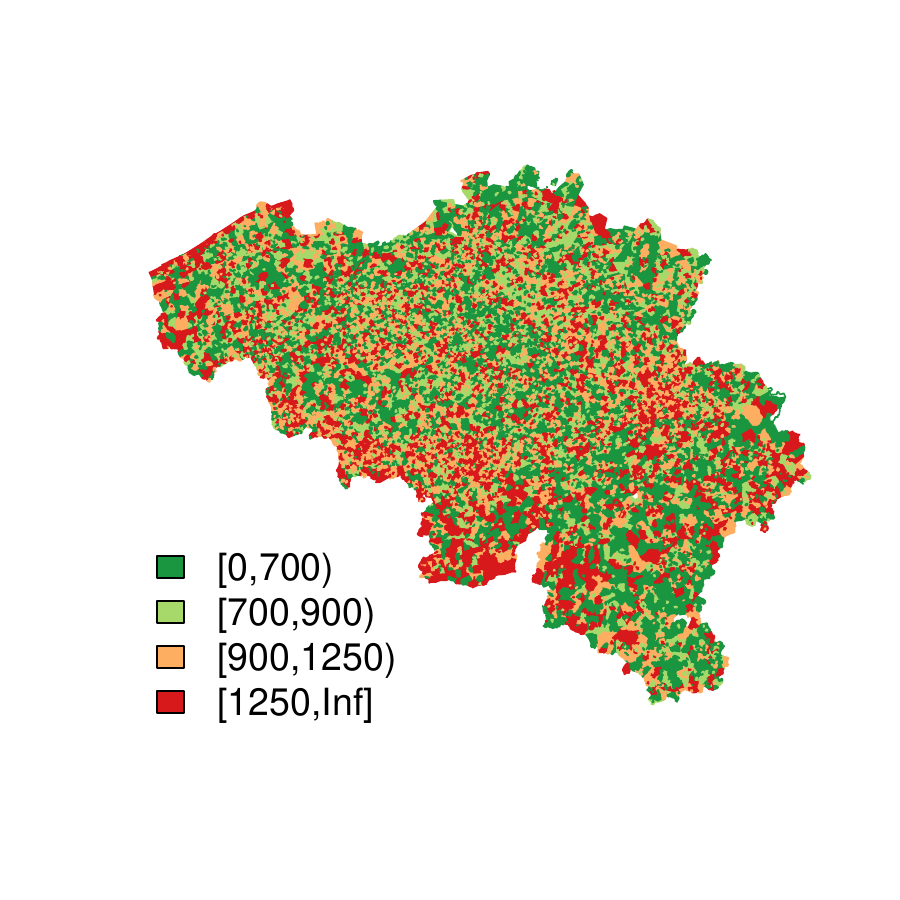}
        \caption{}
        \label{fig:map_sector_Belgium}
    \end{subfigure} 
    \hfill
    \begin{subfigure}[b]{0.4\textwidth}
        \includegraphics[width=\linewidth]{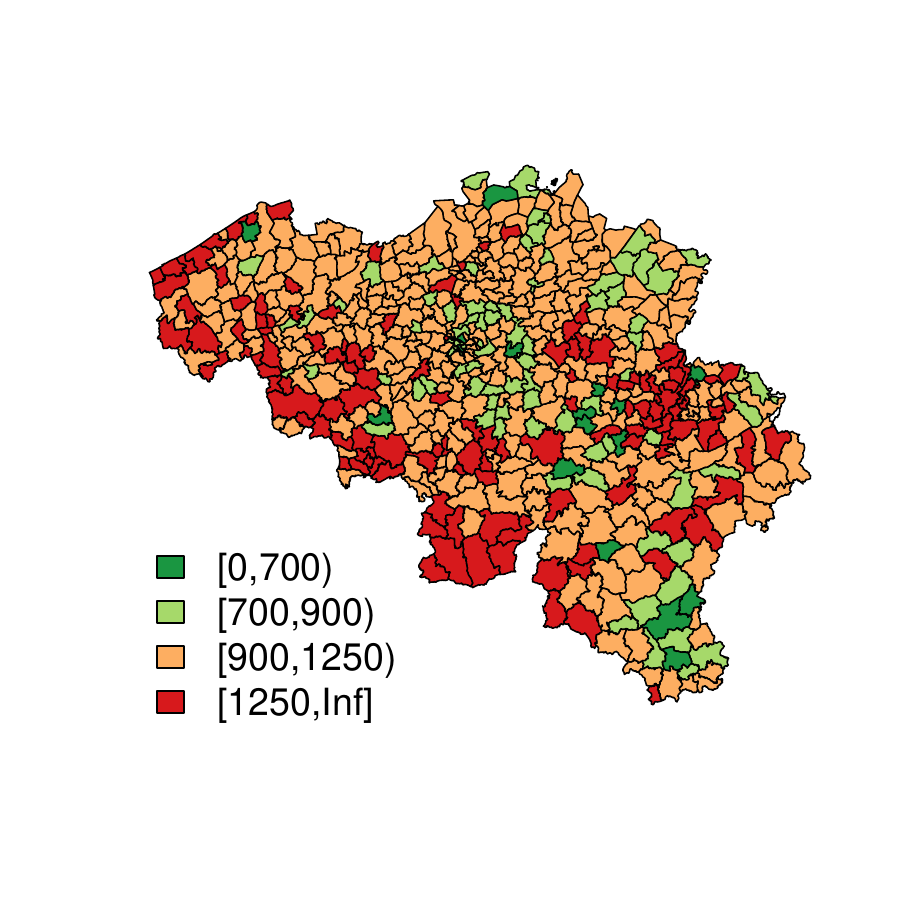}
        \caption{}
        \label{fig:map_area_Belgium}
    \end{subfigure}
\caption{Observed incidences per 100 000 inhabitants: (a) Statistical sector level, (b) Municipality level.}
\label{fig:map_Belgium}
\end{figure}

The dataset provided by Statbel, contains mortality rates from the year 2020, aggregated to the statistical sector level, which is the smallest administrative unit in Belgium, with $19775$ sectors. In this analysis, we aggregated these data to obtain a mortality count for each of the $581$ municipalities in Belgium. A map of the observed counts at the municipality level as well as the statistical sector level is shown in Figure \ref{fig:map_Belgium}. We downloaded a dataset containing information about the gender and educational attainment in each statistical sector, originating from the Belgian census of 2021 from \cite{belgium_stats_sectors}. We used this dataset to calculate the percentage of males, the percentage of people aged over 15 and the percentage of adults who pursued higher education in each statistical sector. These three variables were included linearly in the model. Missing values were inserted using inverse distance weighting interpolation with the \texttt{idw} function in \texttt{gstat} in R. Furthermore, we downloaded a dataset containing the average $\text{PM}_{10}$ pollution level between 1997 and 2022 on a $4 \times 4$ km grid from \cite{pm10_irceline} and a dataset containing the population density on a $1 \times 1$ km grid from \cite{pop_size_statbel}. To allow for possible nonlinear effects, we included all covariates as smooth covariates (with $5$ df). Since our smallest dimension is a $1 \times 1 $ km resolution, we opted to create a $1 \times 1$ km disaggregation raster and extracted all covariate values at every $1 \times 1$ km grid cell (see Figure S15 in the Supplementary Materials). We fitted the model using the population-weighted approximation (SSDAM I). 

\begin{figure}
    \centering
    \includegraphics[width=0.8\linewidth]{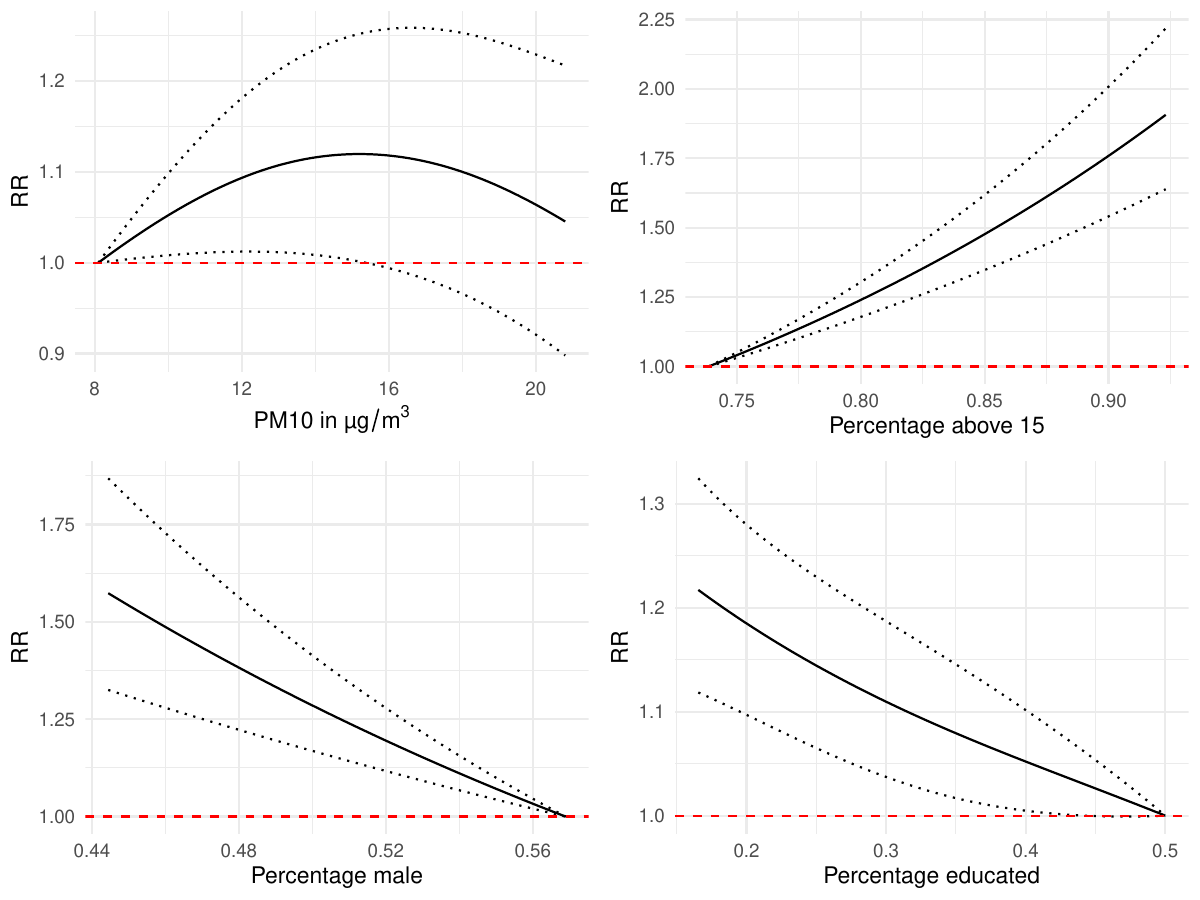}
    \caption{The estimated smooth effect of the four different covariates: $\text{PM}_{10}$ (UL), percentage above 15 (UR), percentage of male (LL) and percentage of educated (LR).}
    \label{fig:smooth_effects}
\end{figure}

The results in Figure \ref{fig:smooth_effects} show that an increase in people over $15$ leads to an increase in expected deaths while an increase in the percentage of males or percentage of people with higher education results in a decreased expected number of deaths. Besides, we find a non-linear increasing RR with higher levels of $\text{PM}_{10}$, up to $15 \ \mu g/m^3$, while the risk is estimated to decrease again once a level of $15 \ \mu g/m^3$ is exceeded. However, this decreasing effect is non-significant because of the high uncertainty. 

\begin{figure}
    \centering
    \begin{subfigure}[b]{0.32\textwidth}
        \includegraphics[width=\linewidth]{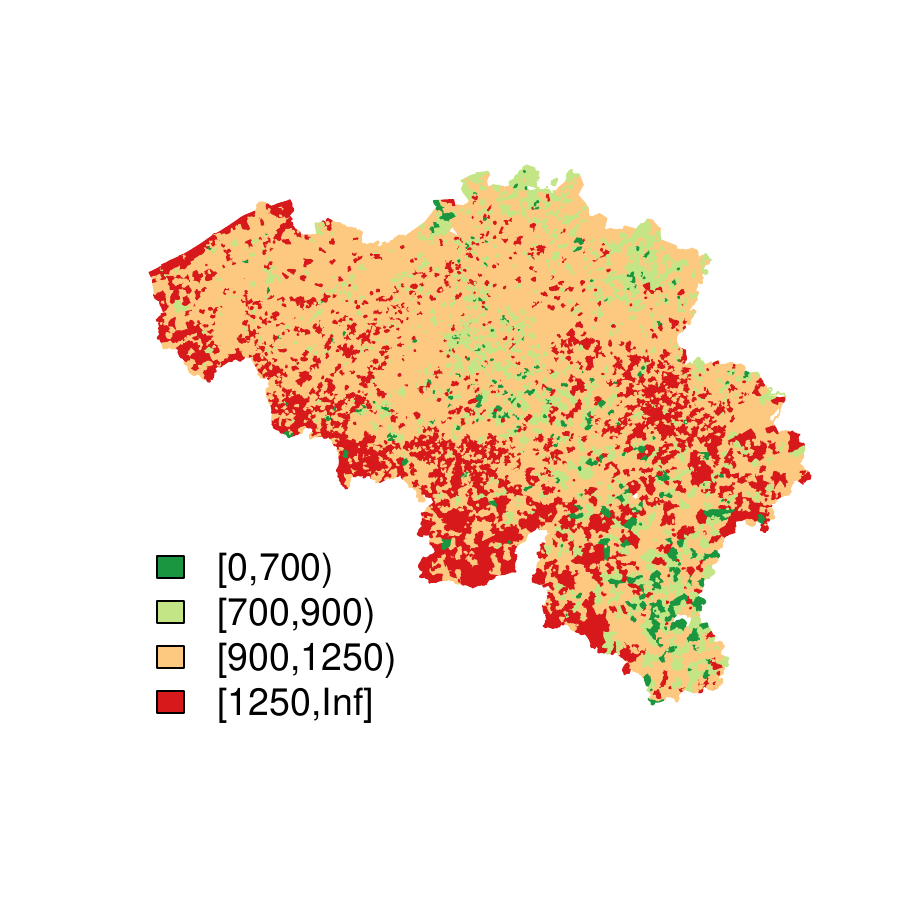}
        \caption{}
        \label{fig:est_sector_approx}
    \end{subfigure} 
    \hfill
    \begin{subfigure}[b]{0.32\textwidth}
        \includegraphics[width=\linewidth]{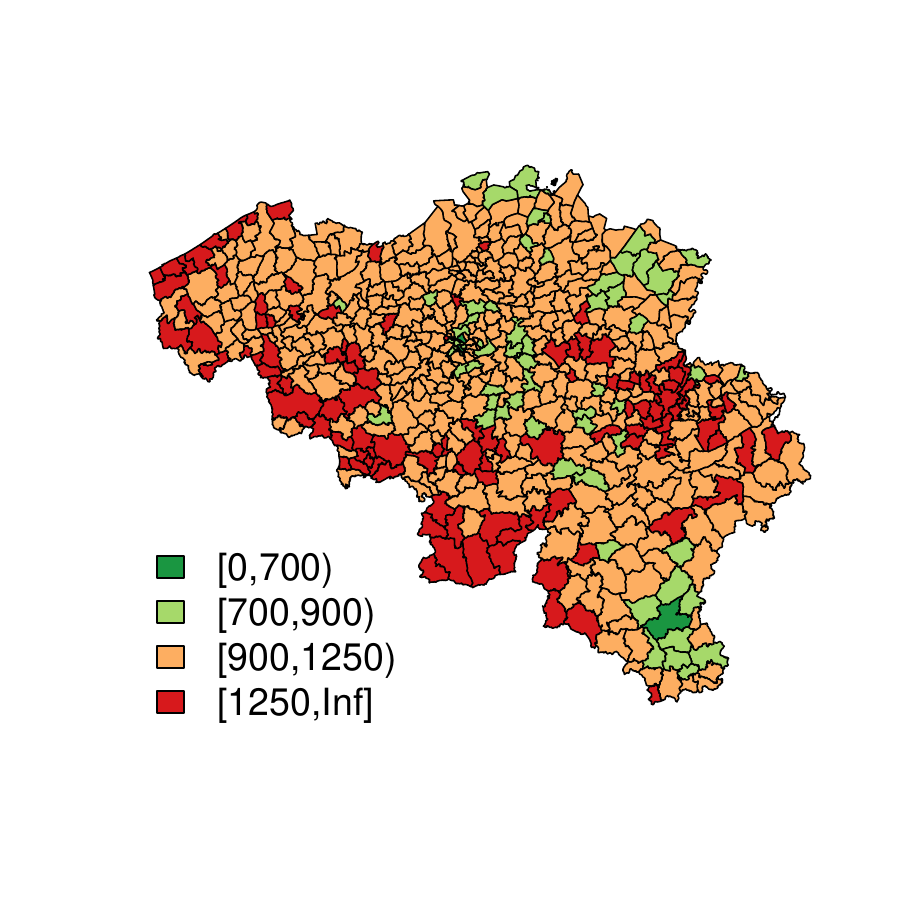}
        \caption{}
        \label{fig:est_area_approx}
    \end{subfigure}
    \begin{subfigure}[b]{0.32\textwidth}
        \includegraphics[width=\linewidth]{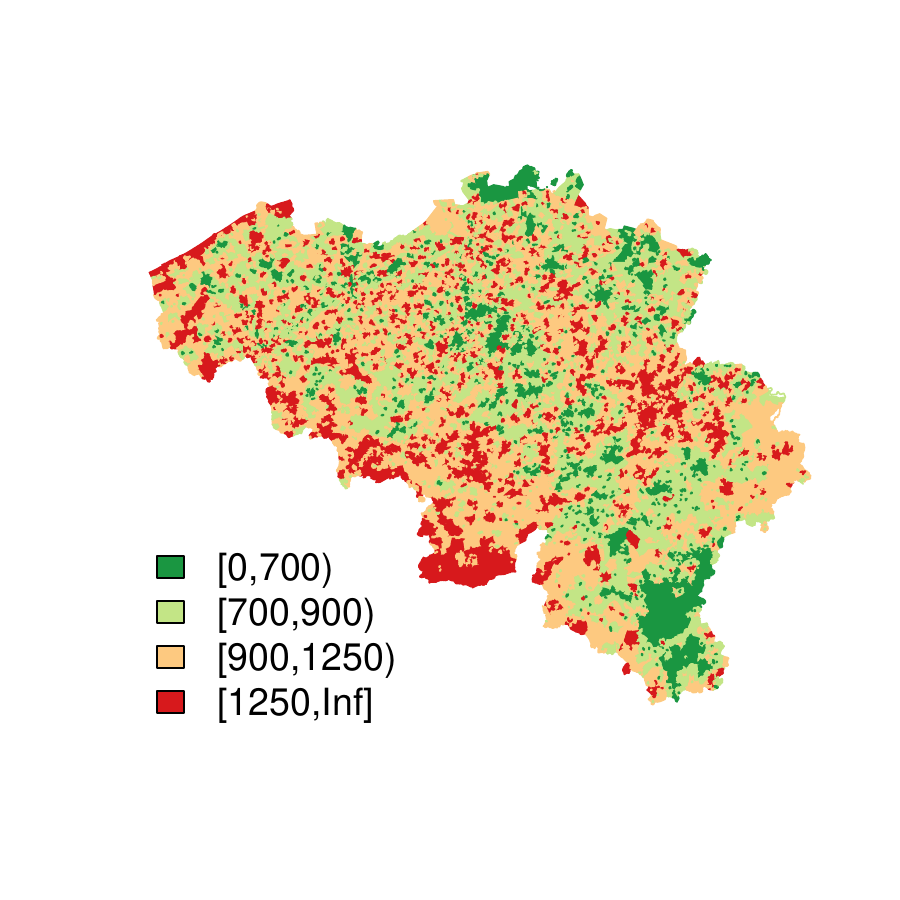}
        \caption{}
        \label{fig:est_INLA}
    \end{subfigure}
\caption{The estimated incidences per 100 000 inhabitants: (a) Statistical sector level, (b) Municipality level and (c) Smoothed incidences at the statistical sector level (using INLA).}
\label{fig:estimates_approx_Belgium}
\end{figure}

Furthermore, we calculated the exceedance probability $r(\boldsymbol{w})>0.01$ on the $1 \times 1$ km grid, where $0.01$ represents the average mortality rate at the statistical sector level (Figure S16 Supplementary Materials). However, a key focus of this analysis is determining whether incidence can be accurately recovered at the statistical sector level. To achieve this, we estimated the incidence at the $1 \times 1$ km raster and aggregated it to the statistical sector level using formula \eqref{eq:approx_matrix}, with a modified $\boldsymbol{A_2}$ matrix, connecting the grid cells to the statistical sectors instead of municipalities. We simulated from the posterior predictive distribution and calculated $95\%$ posterior prediction intervals for the number of deaths in every statistical sector. Doing so, a coverage of $96.94\%$ could be retrieved. In the Supplementary materials, Figure S17 shows the estimated number of deaths, together with the $95\%$ prediction intervals and the true number of deaths in $50$ randomly selected statistical sectors. A map of the estimated incidence at the statistical sector and municipality level can also be found in Figure \ref{fig:est_sector_approx} and \ref{fig:est_area_approx} respectively. At the municipality level, the map of the observed mortality (Figure \ref{fig:map_Belgium}) is similar to the map of the estimated incidences. At the statistical sector level, the differences are larger. However, at a fine spatial scale, observed incidences can be highly variable due to small population sizes and random fluctuations. Therefore, we used INLA to smooth the sector-level incidences, accounting for spatial correlation, in order to reduce the impact of random fluctuations (Figure \ref{fig:est_INLA}). Comparing our estimated incidences at the statistical sector level with the INLA-smoothed incidences, a closer fit can be found. 

We also repeated the analysis using the exact likelihood method (SSDEM) and obtained very similar results. However, the estimation time increased significantly, taking about an hour instead of a few minutes. The results can be found in the Supplementary Materials (Figure S18). Moreover, a comparative figure of the estimated spatial covariance functions is included in the Supplementary Materials as well (Figure S19).

\section{Conclusion}
In this paper, we introduced a novel spatial disaggregation model, allowing for smooth covariate effects through the use of penalized splines. A spline-based low-rank kriging approximation is used to model the spatial component. Laplace approximations are utilized, offering significant computational advantages over the classical MCMC approach. We compared an estimation method based on the exact likelihood (SSDEM) with a method leveraging a spatially discrete approximation (SSDAM). Simulations showed that both methods perform well. Moreover, leveraging approximations results in significant computational gains, particularly important for large datasets. 

Our model can be fitted significantly faster than the SDALGCP model from \cite{Johnson2019}, which makes use of the classical MCMC approach. Furthermore, our methods showed slightly better performance in the presence of large spatial correlation. Simulation studies showed that our method outperforms the \texttt{disaggregation} method, as the latter tends to overestimate the spatial range in the presence of small spatial correlation. Moreover, our methods avoids mesh constructions, which can be tedious, and it allows the inclusion of smooth covariates. While the \texttt{disaggregation} package leverages a template model builder \texttt{TMB} \citep{Kristensen2016}, we derived the analytical calculations underlying the Laplace approximation. We note that our method is slower than the \texttt{disaggregation} package, but leveraging SSDAM still allows the model to be fitted within a few minutes. Lastly, in contrast to the PCLM approach of \cite{Ayma2016}, we obtained estimates of the underlying spatial correlation structure.

Although we think that our method can be very useful for spatial disaggregation purposes, there are some limitations to the approach. First of all, although it offers the advantage of avoiding mesh-construction, the spline-based low-rank kriging approximation still requires the specification of a number of knots, as well as placement of the knots. Simulations showed good performance when using up to $350$ knots, with the number of knots set to the minimum of $350$ and twice the number of areas, across both weak and strong spatial correlation. Nonetheless, this might also depend on the extent of the geographical region. To choose the placement of the knots, we first sample a large amount of locations, reducing them with a space-filling algorithm \citep{Johnson1990,nychka1998} to the specified number of knots. Note that compared to mesh construction, this process is, in our opinion, still more straightforward and less prone to errors. Secondly, the model is relatively complex to estimate and examination of the initial results showed that, especially when using SSDEM, the final estimates of the hyperparameters, and hence also the regression parameters, can be heavily dependent on the initial value of the range parameter. To solve this issue, we decided to run every estimation process multiple times ($25$ by default), starting from different initial values, and then choose the parameters resulting in the highest posterior likelihood. As shown by the results of the simulation study, this approach works well, both for small and strong spatial correlation. Although this method increases computation time, the simulations show that, especially for SSDAM, the model can still be fitted within a few minutes. Moreover, we can see that, although hyperparameter uncertainty is not taken into account, the coverage of the quantities studied in the simulations, approximately reaches the nominal level. 

Hence, despite its limitations, we believe that the introduced model offers a powerful and efficient solution for the increasingly important task of disaggregating count data. Its ability to incorporate penalized smooth covariates, combined with its computational efficiency, makes it a valuable tool not only in epidemiology but also in ecology and environmental science.

\section*{Acknowledgements}
The computational resources and services were provided by the VSC (Flemish Supercomputer Center), funded by the Research Foundation - Flanders (FWO) and the Flemish Government - department EWI. We acknowledge Statbel for providing the Belgian mortality data.  

\section*{Funding}
TN gratefully acknowledges funding by the Research Foundation - Flanders (grant number G0A3M24N).

\section*{Disclosure statement}\label{disclosure-statement}
The authors have declared no competing interest.

\section*{Supplementary materials}
Additional theoretical derivations and figures are available in the Supplementary Materials.

  \bibliography{bibliography.bib}

\end{document}